\documentstyle[epic,eepic,11pt]{article}

\newcommand{\dis}{\displaystyle}

\def\eql(#1){ \begin{equation} \label{#1} 
}
\newcommand{\eq}{\end{equation} }
\begin{document}

\def\b{\beta}
\def\g{\gamma}
\def\G{\Gamma}
\def\Ga(#1,#2){\Gamma^{#1_k #1_{k+1}}_{#2_k #2_{k+1}}}
\def\Prob(#1){P( #1 )}
\def\ket(#1){| #1 \rangle \!\rangle}
\def\bra(#1){ \langle \!\langle #1 |}
\def\vev{``vacuum'' expectation value}
\def\corr{\bf \huge changed}

\begin{flushright}
{SISSA 38/97/EP} \\
{NSF-ITP-97-034} \\
{\today}
\end{flushright}
\vspace*{5mm}

\begin{center}
{ \large \bf  $N$-species Stochastic models with boundaries}
\\ { \large \bf and}
\\ { \large \bf quadratic algebras}\\
[10mm]

\centerline{F. C. ALCARAZ\footnote{Partially supported by 
CNPQ and FAPESP-Brazil}}
\centerline{\hfill}
\centerline{\em Departamento de F\'isica}
\centerline{\em Universidade Federral de S\~ao Carlos}
\centerline{\em 13565-905 S\~ao Carlos, SP, Brazil}
\centerline{\hfill}
\centerline{S. DASMAHAPATRA\footnote{Partially supported by the 
EPSRC grant GRJ 25758}}
\centerline{\hfill}
\centerline{\em Department of Mathematics}
\centerline{\em City University}
\centerline{\em London EC1V 0HB, UK}
\centerline{\hfill}
\centerline{V. RITTENBERG\footnote{Work done under partial support 
of the EC TMR Programme grant FMXR-CT96-0012}}
\centerline{\hfill}
\centerline{\em SISSA, Via Beirut 2-4}
\centerline{\em Trieste I-34014 Italy}
\centerline{and}
\centerline{\em Physikalisches Institut}
\centerline{\em Universit\"at Bonn}
\centerline{\em D-53115 Bonn, Germany}

\end{center}

\newpage 

\pagestyle{empty}

\begin{center}
{\bf ABSTRACT}
\end{center}
\vspace*{5mm}

Stationary probability distributions for 
stochastic processes on linear chains with closed or open 
ends are obtained using the matrix product  Ansatz. The matrices 
are representations of some quadratic algebras. The 
algebras and the types of representations considered depend on the 
boundary conditions. In the language of quantum chains we obtain  the 
ground state of $N$-state quantum chains with free boundary conditions 
or  with non-diagonal boundary terms at one or both ends. In contrast to  
problems involving the Bethe Ansatz, we do not have a general framework
for arbitrary $N$ which when specialized, gives the known results for 
$N=2$; in fact, the $N=2$ and $N>2$ cases appear to be very different.

\vspace*{3mm}

\newpage

\pagestyle{plain}
\renewcommand{\baselinestretch}{1.5}

\section{Introduction}

The aim of this paper is to present in a systematic way the application 
of quadratic algebras to obtain the steady state probability distribution 
of one-dimensional stochastic processes with boundaries. In this section
we first present the problem and describe what is already known about the 
subject giving the relevant references and then proceed by giving the contents 
of the next sections. The list of possible physical applications of our 
results include interface growth \cite{Krug}, 
boundary induced 
phase transitions \cite{DerDomMuk,DomSch,DEHP,Evans-etal},
the dynamics of 
shocks \cite{37} or traffic flow \cite{traf}.

Consider a linear chain with $L$ sites.
On each site $k$ we put a discrete stochastic variable $\beta_k$
taking values from $\{0,1,...,N-1\}$.
For each link $k$ between sites $k$ and $k+1$ we give transition (intensity)
rates
$\Ga(\gamma,\beta)$
giving the probability per unit time for the transition

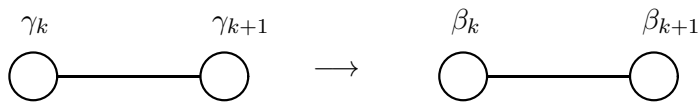
\[
\setlength{\unitlength}{0.0125in}%
\begin{picture}(280,39)(110,690)
\thicklines
\put(120,700){\circle{20}}
\put(200,700){\circle{20}}
\put(300,700){\circle{20}}
\put(380,700){\circle{20}}
\put(130,700){\line( 1, 0){ 60}}
\put(310,700){\line( 1, 0){ 60}}
\put(225,700){$\quad\longrightarrow$}
\put(115,720){$ \gamma_k$}
\put(195,720){$ \gamma_{k+1}$}
\put(295,720){$ \beta_k$}
\put(375,720){$ \beta_{k+1}$}
\end{picture}
\]

These are the {\em bulk rates}.
At the left end of the chain (site 1) we also consider transition rates
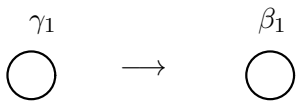
\[
\setlength{\unitlength}{0.0125in}%
\begin{picture}(280,39)(110,690)
\thicklines
\put(200,700){\circle{20}}
\put(300,700){\circle{20}}
\put(225,700){$\quad\longrightarrow$}
\put(195,720){ $\gamma_1$}
\put(295,720){$\beta_1$}
\end{picture}
\]
given by the $N \times N$ matrix $L^{\gamma_1}_{\beta_1}$. 
Similarly,  at the right end of the chain (site $L$) we 
take processes
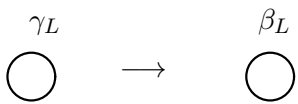
\[
\setlength{\unitlength}{0.0125in}%
\begin{picture}(280,39)(110,690)
\thicklines
\put(200,700){\circle{20}}
\put(300,700){\circle{20}}
\put(225,700){$\quad\longrightarrow$}
\put(195,720){ $\gamma_L$}
\put(295,720){$\beta_L$}
\end{picture}
\]
with rates $R^{\gamma_L}_{\beta_L}$. 
The matrices $L$ and $R$ give the {\em boundary rates}.
We consider Markov processes in continuous time, which implies that 
the rate matrices are {\em intensity} matrices, which have the property 
\cite{Isaacson}
\eql(1.1)
\Ga(\beta,\beta) = -\!\!\! \sum^{N-1}_{
\begin{array}{c}
\scriptstyle \gamma_k, \gamma_{k+1}=0\\
\scriptscriptstyle (\gamma_k, \gamma_{k+1})\neq (\beta_k, \beta_{k+1})
\end{array}} 
\Ga(\beta,\gamma); 
\;
L^{\beta}_{\beta} = - \sum^{N-1}_{
\begin{array}{c}
\scriptstyle \gamma=0\\
\scriptstyle \gamma\neq\beta
\end{array}} L^{\beta}_{\gamma}
;
\;
R^{\beta}_{\beta} = - \sum^{N-1}_{
\begin{array}{c}
\scriptstyle\gamma=0\\
\scriptstyle\gamma\neq\beta
\end{array}} R^{\beta}_{\gamma}.
\eq

Sometimes it will be convenient to write the intensity matrices (for 
example $R$) in an alternative form:
\eql(1.2)
R^{\gamma}_{\beta} = R_{\beta\gamma}
\eq
indicating by the column ($\gamma$) the initial state and by the row ($\beta$) 
the final state. 
Equations (1.1) just state that the sum of the matrix elements on each 
column of an intensity matrix is zero. The time evolution of the probability 
distribution $P(\beta_1,\beta_2,\ldots,\beta_L;t)=:P(\{\beta\};t)$ is 
given by the master equation
\eql(1.3)
\begin{array}{l} \vspace{.2in}
\dis {\partial\over{\partial t}} P(\{\beta\};t) = 
\sum_{k=1}^{L-1} \sum_{\gamma_k, \gamma_{k+1}=0}^{N-1} 
\Ga(\gamma,\beta) 
P(\beta_1, \beta_2, \ldots,\gamma_k, \gamma_{k+1},\ldots, \beta_L ;t)
\\
\quad +
\sum_{\gamma_1=0}^{N-1} L_{\beta_1}^{\gamma_1}P(\gamma_1, 
\beta_2, \ldots,\beta_L;t) 
+
\sum_{\gamma_L=0}^{N-1} R_{\beta_L}^{\gamma_L} P(\beta_1,
\ldots,\beta_{L-1},\gamma_L;t).
\end{array}
\eq
Solving equation (1.3) is equivalent to finding the wave-function 
of an imaginary time Schr\"odinger equation (see \cite{EssRit} 
and references therein for the notations) which is obtained as follows.
We consider an orthonormal system of states
\eql(1.4)
\ket(\beta) \;\; = \;\; \ket(\beta_1 , \ldots \beta_L) \quad;\quad
\bra(\gamma)\cdot\ket(\beta)=\prod_{j=1}^L \delta_{\gamma_j \beta_j},
\eq 
and a basis in the space of $N \times N $ matrices $E^{\alpha \beta}$:
\eql(1.5)
 \left( E^{\alpha \beta} \right)_{\gamma , \delta} 
=
 \delta_{\alpha,\gamma } \delta_{\beta , \delta}. 
\eq
By  $E^{\alpha \beta}_k $  we denote the matrix $E^{\alpha \beta}$ 
acting on the $k^{\mbox{th}}$ site.
The probability distribution $P(\{\beta\} ; t )$ is mapped into a ket state:
\eql(1.6)
\ket(P) \;\; = \;\; \sum_{\{\beta\}}  P(\{\beta\} ; t )  \; \ket(\beta)
\;\; = \;\;
\left( \begin{array}{c} P(0,0,..,0;t) 
\\ P(0,0,..,1;t) \\ \vdots 
\\ P(N\! -\! 1,N\! -\! 1,..,N\! -\! 1;t) 
\end{array}
\right)
\eq
and the master equation (1.3) implies the Schr\"odinger equation 
\eql(1.7)
{\partial\over{\partial t}} \ket(P) 
= \;\; 
-H \; \ket(P)
\eq
where
\eql(1.8)
H = \sum_{k=1}^{L-1} H_{k,k+1} + H_1 + H_L,
\eq
\eql(1.9)
H_{k,k+1} = - \Gamma^{\alpha \beta}_{\gamma \delta} 
E^{\gamma \alpha}_k E^{\delta \beta }_{k+1},
\eq
\eql(1.10)
H_1 = - L^{\alpha}_{\beta} E^{\beta \alpha}_1 \quad \mbox{and} \quad
H_L = - R^{\alpha}_{\beta} E^{\beta \alpha}_L.
\eq
Above, and subsequently, the summation is invoked on {\em any} 
pair of repeated indices (raised, lowered, or mixed).

In this paper we are interested in finding the stationary probability 
distribution $P_s(\{\beta\})$ of the master equation (1.3), i.e. the 
ground state wave function $\ket(P_s)$ of the Hamiltonian (1.8):
\eql(1.11)
H \; \ket(P_s) =0 . 
\eq
Since $H$ is an intensity matrix with positive rates the ground-state energy 
is zero. Unless stated otherwise we consider 
unnormalized probability distributions. The interest in 
knowing the stationary distributions of stochastic processes 
with boundaries is illustrated by the extensive list of references 
on the subject which can be found for example in \cite{Evans-etal}. 
If the bulk rates give a Hamiltonian  with a certain algebra or 
superalgebra 
as symmetry (see ref. \cite{AlcRit} for many examples) the boundary 
terms might break this symmetry.

Recently Krebs and Sandow \cite{KreSan} have proven the following 
remarkable theorem. 
With $\Gamma,L,R$ matrices as above, 
take $N$ matrices $D_{\alpha}$ ($\alpha=0,1,...,N-1$) 
and $N$ matrices $X_{\alpha}$ ($\alpha=0,1,...,N-1$) acting in an 
{\em auxiliary} vector space which satisfy the quadratic algebra relations
\footnote{Observe that the quadratic algebra (1.12) is unusual
since the $X_{\alpha}$'s appear linearly.  (See refs. \cite{Zvyagina}
and \cite{Vershik} for a discussion of quadratic algebras in the
mathematical literature.)}
\eql(1.12)
\Gamma^{\alpha \beta}_{\gamma \delta} D_{\alpha}D_{\beta}=D_{\gamma}
X_{\delta}-X_{\gamma} D_{\delta}, \quad (\gamma, \delta=0,1,\ldots,N-1).
\eq
Consider now a ket state $|0\rangle $ and a bra state $\langle 0|$ 
in the auxiliary space 
(we use the same notation suggesting the ``vacuum'' of quantum mechanics for 
reasons which will be apparent later). If these states are chosen such that
the following conditions are fulfilled:
\eql(1.13)
\langle 0|(X_{\alpha}-L_{\alpha}^{\beta} D_{\beta})=0 \quad \mbox{and}
\quad (X_{\alpha}+R_{\alpha}^{\beta}D_{\beta})|0\rangle =0,
\eq
then
\eql(1.14)
\begin{array}{rcl}
P_s(\{ \beta \} ) &=& \dis \; \langle 0| \; \prod_{k=1}^{L} 
D_{\beta_k} 
|0\rangle \\
&=&\dis \; \langle 0| \; \prod_{k=1}^{L} 
\left( \sum_{\mu_k=0}^{N-1} D_{\mu_k} \delta_{\beta_k, \mu_k} \right) 
|0\rangle 
\end{array}
\eq
is a stationary solution of the master equation (1.3).
Alternatively, let us denote by $u_{\mu_k}$ 
($\mu=0,1,\ldots,N-1$;
$k=1,\ldots,L$) basis vectors in the vector space associated with the
$k^{\scriptstyle th}$ site 
($(u_{\mu_k})_{\nu} =\delta_{\mu,\nu}$ for any $k$) on which 
the matrices $E^{\beta \alpha}_k$ act (see eqs. (1.9-10)).  The ground
state wave-function of the Hamiltonian (1.8) has the expression 
\eql(1.15)
\ket( P_s ) = \langle 0|\prod_{k=1}^L \bigl(\sum_{\mu_k=0}^{N-1} 
D_{\mu_k} u_{\mu_k} \bigr)|0\rangle .
\eq
Actually, the Krebs-Sandow theorem is even stronger; the matrices
$\Gamma^{\alpha \beta}_{\gamma \delta}, R^{\alpha}_{\beta}$ and
$L^{\alpha}_{\beta}$ in (1.8) do not have to be intensity matrices.
If the Hamiltonian has an eigenvalue zero, then the wave function is given
by (1.15).  Moreover, Krebs and Sandow have shown that the algebra
(1.12) exists, giving a representation for the matrices $D_{\alpha}$
and $X_{\alpha}$.  This representation fulfils the second part of 
eq. (1.13) but not the first.

In order to compute the ground state wave-function one needs the 
``vacuum'' expectation values of monomials of the form
\eql(1.16)
\langle 0| D_{\mu_1}^{r_1} D_{\mu_2}^{r_2} \cdots D_{\mu_s}^{r_s} |0\rangle ,
\eq
which are obtained from eqns (1.12) and (1.13).  In order to find the
wave-function of the two-site problem one considers 
``vacuum'' expectation values of monomials of degree
two which are obtained from a system of linear equations.  The solution of
this system is not necessarily unique since the ground state might be
degenerate.  On completing this exercise the $L=3$ \vev s (1.16)
are considered, and so on.  This entire process, however, 
is more complicated than diagonalizing
the Hamiltonian by brute force, and there should be a better way of
solving the problem.  This is indeed possible if the intensity matrices
appearing in the Hamiltonian satisfy certain conditions.  At this point, we
shall mention only what is known in the $N=2$ case.  In reference 
\cite{HinPesSan} Hinrichsen {\em et al}
guessed a four-dimensional representation for the matrices $D_0, D_1, X_0$
and $X_1$ for a specific choice of  intensity matrices and
derived certain concentration profiles.  
In this specific case the problem can also be solved directly since the 
Hamiltonian can be written in terms of free fermions and it was this 
solution which inspired the guess.  This reference is remarkable
since it is the first place where the relations (1.12-16) appear.

We shall now describe the cases where the representation theory is
understood.  Consider the case where the only non-zero bulk rates are
$\Gamma^{10}_{01}$ and $\Gamma^{01}_{10}$ but leave the
surface (boundary) rates
$L_1^0, L_0^1$ and $R_1^0, R_0^1$ arbitrary.  This is the 
asymmetric exclusion model with open boundaries.  One can choose 
$X_1=-X_0=1$ (the c-number $1$) in equations (1.12) and (1.13) and after a
linear transformation $D \rightarrow Y$ which includes constants
related to the boundary rates, bring the relations to the form
\eql(1.17)
[Y_0,Y_1]=\sum_{\gamma, \delta \in\{0,1\}} c^{\gamma \delta} 
Y_{\gamma} Y_{\delta} + \sum_{\gamma\in\{0,1\}}c^{\gamma}Y_{\gamma}
+c_{01}.
\eq
where $[A,B]=AB-BA$ and $c^{\gamma \delta}=c^{\delta \gamma}$.  The
parameters $c^{\gamma \delta}, c^{\gamma}$ and $c_{01}$ 
depend on the bulk and surface rates and we also obtain
\eql(1.18)
\langle 0|Y_1 = 0, \quad Y_0|0\rangle  = 0, \quad \langle 0|0\rangle \neq 0.
\eq
The quadratic algebra with two generators (1.17) is well understood 
\cite{Vershik} and the Fock representations defined by eq. (1.18)
are known \cite{DEHP,EssRit}.  It was in the work of Derrida {\em et al}
that Fock representations of quadratic algebras were used for the first 
time to find ground state wave-functions and to compute 
correlation functions.  As one can notice
from (1.17), (1.18), the \vev s of a monomial of degree $n$ in the
$Y_{\alpha}$'s is determined by the \vev s of a monomial of degree $n-1$
and one of degree $n-2$ in the $Y_{\alpha}$'s.  This implies the existence
of recurrence relations between stationary distribution functions
$P_s(\{\beta\})$ of lattices of different length $L$ (which is also 
the degree of the monomial in question).  The first calculation of
correlation functions in this model were performed using these recurrence
relations without using the algebraic approach \cite{DerDomMuk,DomSch}.

One may try to take more bulk rates than the purely diffusive or hopping
ones just described.  The problem is that (1.12) with $X_{\alpha}$'s chosen
to be c-numbers gives three equations for the two generators $D_0$ and
$D_1$, while one is enough to determine all \vev s as we have just seen.
This implies relations between the rates.  One trivial case corresponds to
one-dimensional representations for the $D_{\alpha}$ (see \cite{EssRit})
when the (connected) correlation functions vanish.  Even two-dimensional
representations don't exist, since the relations on the rates obtained are
incompatible with their positivity \cite{KreSanSim}.  Another possibility
is to not restrict the $X_{\alpha}$'s to be c-numbers, but (say) to take 
\eql(1.19)
X_{\alpha}=A_{\alpha \gamma} D_{\gamma} + x_{\alpha},
\eq
where $A_{\alpha \gamma}$ is an arbitrary $2\times 2$ matrix and
$x_{\alpha}$'s are arbitrary c-numbers.  Using the arbitrariness of the
matrix elements $A_{\alpha\gamma}$, Krebs {\em et al} \cite{KreSanSim}
have shown that it is possible to include more non-zero rates than just
the hopping ones ($\Gamma^{01}_{10}$ and $\Gamma_{01}^{10}$) and get a 
two-dimensional representation of the quadratic algebra, and hence
obtain non-trivial correlation functions.

We would like to mention that the algebraic approach can be used not only in
sequential processes described by the master equation, but also for
parallel dynamics \cite{Hinrichsen,HonPes,Rajewsky-etal}, where the
algebraic problems to be solved are identical.  Also, it is well worth
recalling that matrix-product states first appeared in the literature for
periodic boundary conditions in the work of Hakim and Nadal \cite{HakNad},
and have also appeared in various publications by Fannes, Nachtergale and 
Werner \cite{FanNacWer} who also considered Bethe lattices.  Parallel work
was done by the K\"oln group \cite{Koeln}.  The algebraic problem for
periodic boundary conditions is very different since in this case, the
wave-function is given in terms of traces of monomials and not in terms of
inner products in Fock states.
Quadratic algebras can be used in this case too 
(see Refs.\cite{shocks,34,AHR}).

In the present paper, we address the problem of stochastic processes
involving $N\geq 2$ species.  We shall consider, for the first time, 
not only the case of open
boundary conditions, but also closed or mixed boundary conditions since the
algebraic approach applies to these cases as well.
In section 2, we first review some properties of
intensity matrices which we shall use in the following sections.  In section
3, we consider the problem of the linear chain with closed ends (free
boundary conditions).  The boundary matrices $R$ and $L$ are identically 
zero and we shall consider solutions where $X_{\alpha}=0$.  
This case is not only interesting on its own, but it
represents the natural first step before considering the chain with open
ends where $R$ and $L$ are non-zero.  (The choice $X_{\alpha}=0$
is compatible with eq. (1.13) and will take us to an interesting class 
of quadratic algebras.)  This implies
\eql(1.20)
\Gamma^{\alpha \beta}_{\gamma \delta}D_{\alpha} D_{\beta} =0.
\eq
We shall call a quadratic algebra which contains only quadratic 
terms like in eq.(1.20) a {\em polynomial algebra}. 
(Obviously the algebra defined by eq.(1.17) is not one.)
Equation (1.20) describes relations among monomials of degree two.  The
number of relations depends on the rank of the bulk intensity matrix and
using eq. (1.14), we get the probability distribution for the two-sites
case.  In order to have a solution for more sites, one needs consistency
conditions which imply that the two ways of relating the cubic monomials
$D_{\alpha} D_{\beta} D_{\gamma}$ and $D_{\gamma} D_{\beta} D_{\alpha}$
as depicted in the figure below
\eql(1.21)
\begin{array}{ccccccc}
 & & D_{\alpha} D_{\gamma} D_{\beta} & {\rightarrow} &
D_{\gamma} D_{\alpha} D_{\beta} & & 
\\
 &{\nearrow} &  &  &  & {\searrow} & \\
D_{\alpha} D_{\beta} D_{\gamma} & & & & & &  D_{\gamma} D_{\beta} D_{\alpha} 
\\
 &{\searrow} &  &  &  & {\nearrow} & \\
 & & D_{\beta} D_{\alpha} D_{\gamma} & {\rightarrow} &
D_{\beta} D_{\gamma} D_{\alpha} & & \\
\end{array}
\eq
give the same result.  The requirement of commutativity of the above
diagram imposes constraints on the bulk rates and solves the $L=3$ case.
In all the examples presented in the paper, we have checked that once
the relations (1.21) are satisfied, no supplementary conditions on the
rates arise from quartic (and in some cases) higher degree monomials.
Similar conditions appear in quantum group structures 
\cite{Manin,Sudbery,FTR}
and it is no accident that various quantum planes or superplanes are
solutions of the consistency conditions in some of the cases that we consider. 
Let us observe that for the closed chain case, one does not need to take 
average values 
$\langle 0| \cdots |0\rangle $ in
eq. (1.14); the ground state wave-functions (the ground state is often
degenerate) and therefore the probablity distributions are the 
{\em coefficients of the independent monomials}.  
In section 3 we also answer the following question: if the ground state 
can be obtained 
by the present algebraic methods, is the Hamiltonian integrable?  We shall
show that the answer is in general, no.

In section 4, we consider the case where the left end of the linear chain is
open ($L_{\beta}^{\alpha}\neq 0$) and the right end is closed
($R^{\alpha}_{\beta}=0$).  We again take $X_{\alpha}=0$ but now
we need to add the condition:
\eql(1.22)
L_{\alpha}^{\beta}\langle 0| D_{\beta}=0 \quad (\alpha=0,1,..,N-1)
\eq
to equation (1.20), which comes from eq. (1.13).
This implies new consistency conditions between the bulk and the
left-boundary intensity matrices, which come from the equations
\eql(1.23)
L_{\alpha}^{\beta}\langle 0| D_{\beta} D_{\gamma} = 0.
\eq
The consistency conditions depend on the rank of the $L$ matrix.  Note that
in this case, one has to drop the $|0\rangle $ symbol and retain the $\langle 
0|$ symbol 
in equation (1.14), where the stationary probability distribution is
expressed.  Again there are no Fock representations in this case.

Fock representations appear in section 5, where we consider the linear chain
with both ends open.  This problem is much more difficult than the 
preceding ones.  We therefore limit ourselves to the case $N=3$ only, 
since this case
is complex enough and essentially different from the $N=2$ case.  We also
restrict our attention to the case of diffusion only (in the bulk), {\em
i.e.} the only non-zero bulk rates are
$\Gamma^{\alpha \beta}_{\beta \alpha}$.  We look for representations for the
case where the $X_{\alpha}$'s are c-numbers.  To our knowledge, only one
example of this kind was known up to now \cite{Evans-etal}.  
  It is possible to perform a
linear transformation to bring (1.12) and (1.13) into the form
\eql(1.24)
[Y_{\alpha},Y_{\beta}]=\sum_{\scriptstyle \gamma, \delta \in\{0,1,\ldots,N-1\}}
c^{\gamma \delta}_{\alpha \beta} Y_{\gamma} Y_{\delta} + 
\sum_{\scriptstyle\gamma\in\{0,1,\ldots,N-1\}}c^{\gamma}_{\alpha \beta} 
Y_{\gamma} + c_{\alpha \beta},
\eq
where 
\eql(1.25)
c^{\gamma \delta}_{\alpha \beta}=c^{\delta \gamma}_{\alpha \beta}=
-c^{\gamma \delta}_{\beta \alpha}; \quad 
c^{\gamma}_{\alpha \beta}=-c^{\gamma}_{\beta \alpha};
\quad c_{\alpha \beta}=-c_{\beta \alpha},
\eq
with $N=3$, and the Fock conditions derived from eq.(1.13).  
We review the properties of
the algebra (1.24) and stress that their properties are essentially
different if $N>2$.  We also show that the Fock conditions 
for $N>2$ are in general too numerous.
This leads to a careful separation of cases depending on the rank
of the boundary intensity matrices.

In section 6, we show that if all the minors of the boundary matrices are
non-zero, one can have representations of the quadratic algebra of
dimension at most two  with the bulk and boundary rates lying on 
some algebraic variety.  In section 7, we consider the case in
which only one minor of $L^{\alpha}_{\beta}$ and one of $R^{\alpha}_{\beta}$
are non-zero.  We give the representations of the algebra in this case.  
Other cases (three or more minors non-zero, for example) are not presented 
since the paper is long enough even without them.  In 
Appendix A, we present simple physical processes where the formalism may be
applied.  (The results in the different sections of the paper may be used
for several other applications.)  We consider the problem of spontaneous
symmetry breaking in a two species exclusion model with asymmetric diffusion
proposed by Evans {\em et al} \cite{Evans-etal}.  
More boundary and bulk rates are considered than in the original model. 
This extension might allow a better understanding of the physics of the 
problem.

In section 8, we consider the case in which all the principal
cofactors are zero 
(the boundary intensity matrices have rank one).  We are going to give 
four examples for this case.

The reader who might be tired of this long introduction is invited to go
directly to section 9, where a guide to finding the new results is presented
along with a number of open questions.

\setcounter{equation}{0}
\section{Some properties of intensity matrices}

An $N\times N$ matrix $M$ whose elements $M_{ij}$ are such that 
the sum of all its entries on each column vanishes, {\em i.e.}
\eql(2.1)
\sum_{i=0}^{N-1}M_{ij}=0, \quad (j=0,1,\ldots,N-1),
\eq
is called an {\em I-matrix}.  I-matrices are closed under multiplication 
and form an algebra.

For stochastic problems, one considers I-matrices which have to satisfy 
the further restriction that the off-diagonal elements are real and non 
negative, 
since they are interpreted as probability rates of certain processes.  
Such an $N\times N$ matrix $M$ with real entries $M_{ij}$ is called an 
{\em intensity matrix} if 

\begin{enumerate}

\item all its off-diagonal entries are non-negative, $M_{ij}\geq 0, i\neq j$,

\item and the diagonal elements are negative with 
\eql(2.2)
M_{ii}=-\sum_{i\neq j} M_{ij}.
\eq

\end{enumerate}

For any matrix $A$ with entries $A_{ij}$, if the Laplace row 
expansion  of a determinant is written as
$\mbox{det} A\delta_{ik} = \sum_{j} A_{ij}{\cal A}_{kj}$
where ${\cal A}_{ij}$ is the {\em cofactor} corresponding to the 
$(i,j)^{\scriptstyle th}$
element of $A$.  From this we can infer an important property of any 
zero-column sum matrices; namely their
cofactors for each column are equal in magnitude, and in the 
case of intensity matrices, all the cofactors have the 
same sign.  Let us denote the column-$j$ cofactor
of an intensity matrix $M$ by the corresponding symbol 
in calligraphic font ${\cal M}_j$.

It therefore follows that a system of linear equations
\eql(2.3)
\sum_{n=0}^{N-1} M_{mn} x_n = 0
\eq
has a solution given by
\eql(2.4)
x_n = \xi {\cal M}_n
\eq
for any constant (independent of $n$) $\xi$ if the rank of $M$ is $N-1$.

Consider two intensity matrices $F$ and $G$, and their sum
\eql(2.5)
E=F+G,
\eq
which is also an intensity matrix.  The following identity is true for
$N=1,2$ and $3$ only:
\eql(2.6)
\sum_{m=0}^{N-1}(F_{nm}-G_{nm}){\cal E}_m = (N-1) \sum_{m=0}^{N-1}
(F_{nm} {\cal G}_m - G_{nm} {\cal F}_m),
\eq
where as noted earlier, the calligraphic ${\cal E, F, G}$ with subscript
$m$ denotes the cofactors of $E, F, G$ for the $m^{\scriptstyle th}$
column.  The last identity is useful for the $3$ species case.
Since we are going to use them often in the next few sections, we give the
expressions for the three distinct cofactors of a 
$3\times 3$ intensity matrix
$F$ with matrix elements $F_n^m$:
\eql(2.7)
\begin{array}{ccc}
{\cal F}_0& =&  F^1_0 F^2_1 + F^2_0 F^1_0 + F^2_0 F^1_2,\\
{\cal F}_1& =&  F^0_1 F^2_0 + F^2_1 F^0_1 + F^2_1 F^0_2, \\
{\cal F}_2& =&  F^0_2 F^1_0 + F^1_2 F^0_2 + F^1_2 F^0_1.
\end{array}
\eq
%
% para

\setcounter{equation}{0}
\section{Steady states for a linear chain with closed ends
(Ground states for quantum spin chains with free boundary conditions)}

As mentioned in the introduction, we consider ground state wave-functions of
the form
\eql(3.1)
\ket( P_s )=\prod_{k=1}^L 
\big( \sum_{\mu_k=0}^{N-1} D_{\mu_k} u_{\mu_k}
\bigr),
\eq
where the matrices $D_{\mu}$ satisfy equations (1.20) and the consistency
conditions (1.21). The type of wave-functions we get depends on the rank of
the bulk intensity matrix $\Gamma^{\alpha \beta}_{\gamma \delta}$.  Let us
first consider the case where the rank is 
the maximum possible, i.e. $N^2-1$, and all principal minors are
non-zero.  We denote by
${\cal G}_{\alpha \beta}$ the cofactors of $\Gamma^{\alpha \beta}_{\gamma
\delta}$ (see section 2).  Using equation (2.4) we get 
\eql(3.2)
D_{\alpha} D_{\beta} = {{\cal G}_{\alpha \beta} \over {\cal G}_{\gamma
\delta}} D_{\gamma} D_{\delta}.
\eq

There are two ways in which the cubic monomial 
$D_{\alpha} D_{\beta} D_{\gamma}$ can be related to 
$D_{\mu} D_{\rho} D_{\sigma}$:
\eql(3.3)
\begin{array}{cccccc}
(i) & D_{\alpha} D_{\beta} D_{\gamma} & = &\dis {{\cal G}_{\alpha \beta} 
\over {\cal G}_{\mu \nu}} D_{\mu} D_{\nu} D_{\gamma} & = & \dis 
{{\cal G}_{\alpha \beta} \over {\cal G}_{\mu \nu}}
{{\cal G}_{\nu \gamma} \over {\cal G}_{\rho \sigma}}
D_{\mu} D_{\rho} D_{\sigma} \\
&&&&& \\
(ii) & D_{\alpha} D_{\beta} D_{\gamma} & = & \dis {{\cal G}_{\beta \gamma} 
\over {\cal G}_{\nu \sigma}} D_{\alpha} D_{\nu} D_{\sigma} & = & 
\dis {{\cal G}_{\beta \gamma} \over {\cal G}_{\nu \sigma}}
{{\cal G}_{\alpha \nu} \over {\cal G}_{\mu \rho}}
D_{\mu} D_{\rho} D_{\sigma} 
\end{array}
\eq
Comparing (i) and (ii) in equation (3.3) (and setting 
$\rho=\nu$)
we get:
\eql(3.4)
{\cal G}_{\alpha \beta}{\cal G}_{\nu \gamma}
= {\cal G}_{\alpha \nu}{\cal G}_{\beta \gamma}
\eq
and hence the condition on the cofactors (on setting $\nu=\alpha$
and $\beta=\gamma$) is
\eql(3.5)
{\cal G}_{\alpha \beta} = \pm \sqrt{ 
{\cal G}_{\alpha \alpha}{\cal G}_{\beta \beta}} = 
{\cal G}_{\beta \alpha}.
\eq
The $+$($-$) is taken if $N$ is odd (even).  This implies that the
$D_{\alpha}$'s can be taken as c-numbers:
\eql(3.6)
D_{\alpha}=\xi \sqrt{{\cal G}_{\alpha \alpha}},
\eq
where $\xi$ is a constant.  The ground state is unique and the 
(connected) correlation functions manifestly zero.

The whole structure of the possible solutions for the ground state
wave-function depends on the rank of the bulk intensity matrix
$\Gamma_{\gamma \delta}^{\alpha \beta}$.  We shall restrict 
ourselves to the cases $N=2,3$.  Before proceeding further, we 
would like to consider some questions about integrability.  We 
are going to find that for some intensity matrices we can find 
ground-state wave-functions. In general the ground is degenerate,
which might imply the existence of some symmetry in the Hamiltonian.
What is this symmetry?  On the other hand, one can also 
ask the (possibly related) question, if the ground states are obtained
from polynomial algebras, is the Hamiltonian integrable?  
In general, the answer is negative.  Exact integrability requires
more than the exact expression for the ground-state wavefunction.

In certain cases, the $e_k$ defined by
\eql(3.7)
e_k:=H_{k,k+1},
\eq
where $H_{k,k+1}$ is the Hamiltonian density given in equation (1.9),
satisfy the relations of a Hecke algebra:
\eql(3.8)
\begin{array}{rcl}
e_k e_{k\pm 1} e_k - e_k & = &e_{k\pm 1} e_k e_{k\pm 1} - e_{k\pm 1} \\
\left[e_k , e_j\right] & = & 0, \quad \mbox{for} \; |k-j|\geq 2\\
e_k^2 & = &(q+q^{-1}) e_k.
\end{array}
\eq
In this case, one can find a spectral parameter dependent 
solution $\check R(u)$ of the Yang-Baxter equation via "Baxterization"
\cite{Jones,Baxter}.  One can check that the generators $e_k$ satisfy
further relations among themselves which define various quotients of the 
Hecke algebra \cite{Martin}.
For our purposes, it is sufficient to state that the 
$H_{k,k+1}$ belong to  such quotients that 
are classified by a pair of natural numbers $(P,M)$.  For each
quotient labelled by $(P,M)$ the Hamiltonian whose density is
given by the appropriate $e_k$ defines a spin chain \cite{Perk-Schultz}
with $U_q su(P|M)$ (which is
the Schur-Weyl dual of the $(P,M)$ Hecke quotient) as its 
symmetry algebra.  (For further details, see \cite{AlcRit}.)

A more general criterion for integrability at the level of 
Hamiltonian densities was introduced by Reshetikhin \cite{Res}
for the case of $R$-matrices whose dependence on the spectral
parameter is of difference type.  By expanding the $R$ and 
row transfer matrices in powers of the spectral parameter, 
it was shown to be necessary that 
the Hamiltonian densities $e_k$ satisfy the following relation
\eql(3.9)
[e_k+e_{k+1},[e_k,e_{k+1}]]=W_k-W_{k+1},
\eq
where $W_k$ is a tensor product of identity matrices at all
sites on the chain and an arbitrary matrix at the 
$k^{\scriptstyle th}$ site.

\vskip 4pt

\noindent $\bullet\underline{N=2}$

\vskip 4pt

We now consider the $N=2$ case.  We shall on occasion refer to the
state ``$1$'' as that denoting the presence of a particle 
and the state ``$0$'' to its absence.
If all the minors are non-zero,
and the intensity matrix is of maximal possible rank ({\em i.e.} $3$)
equation (3.6) applies.  
We consider the cases where the rank is less than $3$.  
Let us first consider the case 
\eql(3.10)
\Gamma_{10}^{00}=\Gamma_{01}^{00}=\Gamma_{11}^{00}=
\Gamma_{00}^{01}=\Gamma_{00}^{10}=\Gamma_{00}^{11}=0,  
\eq
where the bulk intensity matrix effectively reduces (from a
$4\times 4$) to a $3\times 3$ matrix with rows and columns labelled by
$(1,0)$, $(0,1)$ and $(1,1)$.  Let ${\cal G}_{1,0}'$, ${\cal G}_{0,1}'$
and ${\cal G}_{1,1}'$ be the cofactors of this new intensity matrix.  
In order to have a polynomial algebra we must have the condition
\eql(3.11)
{\cal G}_{1,0}' = {\cal G}_{0,1}',
\eq
and the algebra is 
\eql(3.12)
D_1 D_0 = D_0 D_1 = {{\cal G}_{1,0}' \over {\cal G}_{1,1}'} D_1^2.
\eq

The ground state for $L$ sites is now doubly degenerate.  One state
corresponds to $D_0^L$ and the other contains all the  
monomials which have at least one $D_1$, and can be brought 
by eq.(3.12) to be proportional to $D_1^L$.  

Next we consider the situation where we have only diffusion of particles
and their annihilation if two particles are on neighbouring sites.
This is attained by choosing 
the following rates to vanish (the rest being non-zero):
\eql(3.13)
\Gamma_{10}^{00}=\Gamma_{01}^{00}=\Gamma_{11}^{00}=\Gamma_{00}^{01}=
\Gamma_{00}^{10}=\Gamma_{11}^{01}=\Gamma_{11}^{10}=0.  
\eq
In this case the algebra is
\eql(3.14)
D_0 D_1 = q^2 D_1 D_0; 
\quad D_1^2 = 0,
\eq
where
\eql(3.15)
q=\sqrt{ \Gamma_{01}^{10} \over \Gamma_{10}^{01}}.
\eq
If we choose the time scale such that $\Gamma^{10}_{01}\Gamma^{01}_{10}=1$,
the set of $H_{k,k+1}$ satisfies the relations for the $(1,1)$
quotient of the Hecke algebras centralizing $U_q(su(1|1))$.  
The ground state for $L$ sites is doubly degenerate corresponding
to the words $D_0^L$ and $D_1 D_0^{L-1}$.

\def\Ga(#1,#2){\Gamma^{#1}_{#2}}
Yet another example of an intensity matrix 
which gives rise to a quadratic algebra is
\eql(3.16)
\left(\begin{array}{cccc}
-\Ga(00,10) & 0 & \Ga(10,00) & 0 \\
0 & -\Ga(01,11) & 0 & \Ga(11,01) \\
\Ga(00,10) & 0 & -\Ga(10,00) & 0 \\
0 & \Ga(01,11) & 0 & -\Ga(11,01)
\end{array}\right)
\eq
which gives the following quadratic algebra:
\eql(3.17)
D_0 D_1 = {\dis 1\over q^{2}} D_1^2 \quad \mbox{and} 
\quad D_1 D_0 = q^{2} D_0^2,\quad \mbox{where}\quad
q^2=\dis {\Ga(00,10)\over \Ga(10,00)}={\Ga(01,11)\over\Ga(11,01)}
\eq
after solving for the associativity constraints coming from the
cubic terms.

Yet another possibility is to set more rates to $0$, in addition to those
which led to equation (3.14).  Upon setting $\Gamma_{00}^{11}$,
$\Gamma_{01}^{11}$ and $\Gamma_{10}^{11}$ to zero as well, 
which means we are only considering diffusion, we are led to
\eql(3.18)
D_0 D_1 = {\Gamma^{10}_{01} \over \Gamma^{01}_{10}} D_1 D_0.
\eq
The ground state is now $(L+1)$ times degenerate -- each wave-function has a
given number of $D_1$'s and the coefficients of the tensor products of the
$u_k^{(\mu)}$'s in eq.(3.1) correspond to the $q$-deformation of the 
symmetrizer corresponding to the Young 
diagram with one row and $L$ boxes \cite{AlcRit}, with
$q=\sqrt{\Gamma^{10}_{01}/\Gamma_{10}^{01}}$.
The quadratic relation 
\eql(3.19)
D_0 D_1 = 0
\eq
corresponds not only to the case where $\Gamma^{01}_{10}$ is the
only non-zero rate, but more generally $\Gamma^{01}_{\alpha \beta}$ is
non-zero for all $\alpha, \beta$.  Again the ground state is $(L+1)$ times
degenerate.  Since our intention is to generalize the problem to $N>2$, it
is useful to present visually the non-zero matrix elements which correspond
to the different quadratic relations.  The rows and columns below are
labelled $(0,0),(0,1),(1,0)$ and $(1,1)$ consecutively.

\def\Ga(#1,#2){\Gamma^{#1}_{#2}}

$$\left(\begin{array}{cccc}
\Box & \Box & \Box & \Box \\
\Box & \Box & \Box & \Box \\
\Box & \Box & \Box & \Box \\
\Box & \Box & \Box & \Box
\end{array}\right) \quad \mbox{eq.}(3.6)\quad
\left(\begin{array}{cccc}
\cdot & \cdot & \cdot & \cdot \\
\cdot & \Box & \Box & \Box \\
\cdot & \Box & \Box & \Box \\
\cdot & \Box & \Box & \Box
\end{array}\right)\quad\mbox{eq.}(3.12)$$

$$\left(\begin{array}{cccc}
\cdot & \cdot & \cdot & \Box \\
\cdot & \Box & \Box & \Box \\
\cdot & \Box & \Box & \Box \\
\cdot & \cdot & \cdot & \Box 
\end{array}\right) \quad\mbox{eq.}(3.14)\quad
\left(\begin{array}{cccc}
\Box & \cdot & \Box & \cdot \\
\cdot & \Box & \cdot & \Box \\
\Box & \cdot & \Box & \cdot \\
\cdot & \Box & \cdot & \Box
\end{array}\right) \quad\mbox{eq.}(3.17)$$

$$\left(\begin{array}{cccc}
\cdot & \cdot & \cdot & \cdot \\
\cdot & \Box & \Box & \cdot \\
\cdot & \Box & \Box & \cdot \\
\cdot & \cdot & \cdot & \cdot
\end{array}\right) \quad\mbox{eq.}(3.18) \quad
\left(\begin{array}{cccc}
\cdot & \Box & \cdot & \cdot\\
\cdot & \Box & \cdot & \cdot\\
\cdot & \Box & \cdot & \cdot\\
\cdot & \Box & \cdot & \cdot
\end{array}\right) \quad\mbox{eq.}(3.19)$$
We could not obtain algebras with non-scalar $D_0$ and $D_1$ for any 
other choice of locations in the intensity matrix for the non-zero 
rates than those depicted above.

\vskip 4pt

\noindent $\bullet\underline{N=3}$

\vskip 4pt

We now consider the $N=3$ case.  We are not going to list all the possible
quadratic relations for the two-site problem which are compatible with the
$3$-site problem ({\em i.e.} the cubic relations are compatible with the
associative application of the quadratic relations)
but list a few example which do.  
We shall, in some cases, interpret the state variables $1,2,0$ as those of
two species of particles and holes respectively.
First, we consider
the case in which the two-site bulk intensity matrix decouples into three
intensity matrices (the other rates being zero):
$$\left(\begin{array}{cc}
-\Gamma_{10}^{01} &  \Gamma_{01}^{10}\\
\Gamma_{10}^{01} &  -\Gamma_{01}^{10}
\end{array}\right) \quad 
\left(\begin{array}{cc}
-\Gamma_{20}^{02} &  \Gamma_{02}^{20}\\
\Gamma_{20}^{02} &  -\Gamma_{02}^{20}
\end{array}\right),$$
and
\eql(3.20)
\left(
\begin{array}{ccc}
-(\Gamma_{21}^{00}+\Gamma_{12}^{00})& \Gamma_{00}^{12} &  \Gamma_{00}^{21}\\
\Gamma_{12}^{00} & -(\Gamma_{00}^{12}+\Gamma_{21}^{12})   &  \Gamma_{12}^{21}\\
\Gamma_{21}^{00} & \Gamma_{21}^{12} & -(\Gamma_{00}^{21}+\Gamma_{12}^{21})
\end{array}
\right).
\eq
We denote by ${\cal J}_{00}$, ${\cal J}_{12}$ and ${\cal J}_{21}$ 
the cofactors
of the $3\times 3$ intensity matrix.  The quadratic relations obtained are 
\eql(3.21)
\begin{array}{cc}
D_0 D_1 =  \gamma D_1 D_0 & D_2 D_0  =  \delta D_0 D_2, \\
D_1D_2  = \beta D_0^2 & \alpha D_2D_1 = \beta D_0^2.
\end{array}
\eq
The cubic relations obtained by considering words in the $D$'s of degree
$3$ are compatible with the quadratic relations provided
\eql(3.22)
\alpha={{\cal J}_{12} \over {\cal J}_{21}}=\gamma^{-2};\quad
\beta={{\cal J}_{12} \over {\cal J}_{00}};\quad
\gamma=\delta= {\Gamma_{01}^{10} \over  \Gamma_{10}^{01}} = 
{\Gamma_{20}^{02} \over\Gamma_{02}^{20}}. 
\eq
There are $(2L+1)$ wave-functions with energy eigenvalue zero.  They are
given in terms of the monomials 
\eql(3.23)
D_0^{L-m} D_1^m, \quad D_0^{L-n} D_2^n.
\eq
This gives an explanation as to why it was possible to calculate 
the probability distribution exactly in \cite{alcaraz} for the 
particular case where interchange 
of particles  is forbidden ($\Gamma_{21}^{12} = \Gamma_{12}^{21} = 0$).
Interpreting the matrix entries in eq. (3.20) as transition rates 
involving two species of particles $1$ and $2$ we notice that the 
difference in the number of particles of types $1$ and $2$ is 
conserved under this stochastic dynamics.  

We now consider the intensity matrix of the block diagonal form
\eql(3.24)
\left(\begin{array}{ccccccccc}
\Ga(00,00) & \Ga(01,00) & \Ga(02,00) & 
0 & 0 & 0 & 0 & 0 & 0 \\
\Ga(00,01) & \Ga(01,01) & \Ga(02,01) & 
0 & 0 & 0 & 0 & 0 & 0 \\
\Ga(00,02) & \Ga(01,02) & \Ga(02,02) & 
0 & 0 & 0 & 0 & 0 & 0 \\
0 & 0 & 0 & \Ga(10,10) & \Ga(11,10) & \Ga(12,10) & 
0 & 0 & 0 \\
0 & 0 & 0 & \Ga(10,11) & \Ga(11,11) & \Ga(12,11) & 
0 & 0 & 0 \\
0 & 0 & 0 & \Ga(10,12) & \Ga(11,12) & \Ga(12,12) & 
0 & 0 & 0 \\
0 & 0 & 0 & 0 & 0 & 0 & \Ga(20,20) & \Ga(21,20) & \Ga(22,20) \\
0 & 0 & 0 & 0 & 0 & 0 & \Ga(20,21) & \Ga(21,21) & \Ga(22,21) \\
0 & 0 & 0 & 0 & 0 & 0 & \Ga(20,22) & \Ga(21,22) & \Ga(22,22)
\end{array}\right), 
\eq
and we denote the cofactors of the $3\times 3$ block 
containing $\Ga(ii,ii)$ by ${\cal G}^{\alpha}_{(i)}$, 
where $\alpha=0,1,2$ labels the $(\alpha+1)^{\scriptstyle th}$
column within each block.
The algebra is defined by the six relations 
\eql(3.25)
D_{\alpha} D_{\beta} = {f_{\beta}\over f_{\alpha}} D_{\alpha}^2,
\eq
where
\eql(3.26) 
{f_1\over f_0}={{\cal G}_{(0)}^1 \over {\cal G}_{(0)}^0} 
= {{\cal G}_{(1)}^1 \over {\cal G}_{(1)}^0};\quad 
{f_0\over f_2}={{\cal G}_{(2)}^0 \over {\cal G}_{(2)}^2} 
= {{\cal G}_{(0)}^0 \over {\cal G}_{(2)}^0}; 
\eq

\eql(3.27)
{f_2\over f_1}={{\cal G}_{(1)}^2 \over {\cal G}_{(1)}^1} 
= {{\cal G}_{(2)}^2 \over {\cal G}_{(2)}^1};\quad 
{\cal G}_{(0)}^0 {\cal G}_{(1)}^1 {\cal G}_{(2)}^2 
= {\cal G}_{(0)}^1 {\cal G}_{(1)}^2 {\cal G}_{(2)}^0.
\eq
The ground state is $3$-fold degenerate, the independent words
being $D_0^L$, $D_1^L$ and $D_2^L$. This degeneracy arises since the 
type of particle located at the first site $(0,1,2)$ is unchanged
by the stochastic dynamics.

An interesting example with $4$ quadratic relations is
\eql(3.28)
\begin{array}{ccc}
\Gamma^{01}_{10} D_0 D_1 &=& \Gamma^{10}_{01} D_1 D_0 \\
\Gamma^{12}_{21} D_1 D_2 &=& \Gamma^{21}_{12} D_2 D_1 \\
\Gamma^{20}_{02} D_2 D_0 &=& \Gamma^{02}_{20} D_0 D_2 \\
D_1^2 &=& 0.
\end{array}
\eq
{} These relations are obtained if the only non-zero rates are those 
that appear in (3.28) and $\Gamma^{11}_{\alpha \beta}$, ($\alpha,\beta
\in\{0,1\}$).  In order to find the 
degeneracy of the ground state, notice that if we set 
\eql(3.29)
\Ga(\alpha\beta,\beta\alpha)={1\over\Ga(\beta\alpha,\alpha\beta)},
\;(\alpha\neq\beta)\quad \mbox{and}\quad  {\Ga(01,10)\over\Ga(10,01)}=
{\Ga(02,20)\over\Ga(20,02)}={\Ga(12,21)\over\Ga(21,12)}=q^2
\eq
we obtain the $(2,1)$ quotient of the Hecke algebra \cite{AlcRit}, 
and by Schur-Weyl duality, the Hamiltonian is $U_q(su(2|1))$ symmetric.  
The ground state wave-functions are proportional to the monomials 
symmetrized using the relations (3.28,29) and are therefore given
by the representation of the superalgebra which corresponds to a Young 
diagram with $L$ boxes in one row \cite{MarRit}.  

A physically important example is the case in which we 
consider hopping and interchange of particles; we set the rates 
$\Gamma^{11}_{\alpha \beta}$ to zero in the previous example
to get the three relations
\eql(3.30)
D_0 D_1={\Gamma^{10}_{01}\over \Gamma^{01}_{10}} D_1 D_0,\; 
D_1 D_2={\Gamma^{21}_{12}\over \Gamma^{12}_{21}} D_2 D_1,\;
D_2 D_0={\Gamma^{02}_{20}\over \Gamma^{20}_{02}} D_0 D_2. 
\eq
We have checked whether the Hamiltonian thus defined satisfies the 
necessary condition as prescribled by Reshetikhin (eq. (3.9))
for being integrable.  We found that in order to be so, the
rates must satisfy the relations (3.29) and we get the $(3,0)$ 
quotient of the Hecke algebra and the Hamiltonian is $U_q(su(3))$
symmetric.  This result is important because it shows that although the
ground state of the Hamiltonian can be obtained by algebraic means, the
spectrum of the Hamiltonian cannot, unless supplementary
conditions are satisfied.  We have also checked other cases
with the same result.  If the relations (3.29) are satisfied, 
the ground states again give the
representation of $U_q(su(3))$ which correspond to the Young diagram with
$L$ boxes in one row \cite{MarRit}.

We can also consider the case where the 
 only rates that are non-zero are
$\Gamma^{01}_{10}$, $\Gamma_{01}^{10}$, $\Gamma^{02}_{20}$
and $\Gamma^{20}_{02}$.  The only two relations that survive are
\eql(3.31)
\Gamma^{01}_{10} D_0 D_1 = \Gamma^{10}_{01} D_1 D_0 \quad \mbox{and}
\quad \Gamma^{02}_{20} D_0 D_2 = \Gamma^{20}_{02} D_2 D_0,
\eq
and the degeneracies are huge since any sequence of the
state indices $1$ and $2$ is invariant under the dynamics
and can be computed using the results of ref.\cite{Alc-etal}.

Finally, one can have only one relation 
\eql(3.32)
D_i D_j =0,
\eq
if the only non-zero rates are $\Gamma^{i,j}_{\alpha,\beta}$ for 
any $\alpha$, $\beta$. 

Since we have only looked at particular cases, it is most likely that 
we have not exhausted all possible quadratic algebras arising from
$9\times 9$ intensity matrices.  We have not been able to find other 
examples which give non-trivial (non-scalar) representations 
for the relations emerging from a choice of intensity matrix.  The 
rules of this game are however very simple and can be applied 
by the reader to find other examples.  In the $9\times 9$ intensity  
matrix, one takes several diagonal blocks of smaller intensity 
matrices with non-zero rates and columns of possibly non-zero rates
of the form depicted below.

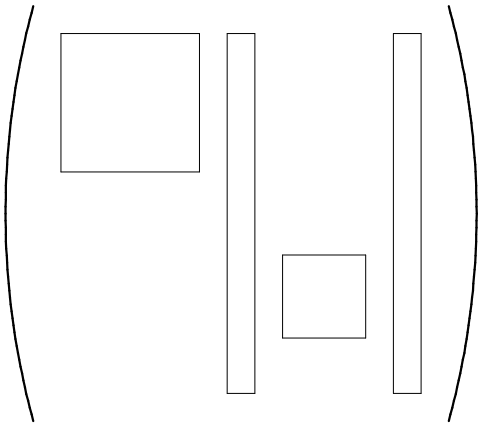
\centerline{
\setlength{\unitlength}{0.00725in}
\begin{picture}(343,315)(0,-10)
\thicklines
\drawline(22.000,300.000)(20.690,295.113)(19.424,290.214)
	(18.202,285.305)(17.023,280.384)(15.887,275.454)
	(14.796,270.513)(13.748,265.564)(12.745,260.605)
	(11.785,255.637)(10.870,250.661)(9.998,245.677)
	(9.171,240.685)(8.389,235.687)(7.650,230.681)
	(6.957,225.670)(6.307,220.652)(5.702,215.629)
	(5.142,210.600)(4.627,205.567)(4.156,200.529)
	(3.730,195.488)(3.349,190.443)(3.012,185.394)
	(2.720,180.343)(2.473,175.290)(2.271,170.234)
	(2.114,165.177)(2.002,160.119)(1.935,155.060)
	(1.912,150.000)(1.935,144.940)(2.002,139.881)
	(2.114,134.823)(2.271,129.766)(2.473,124.710)
	(2.720,119.657)(3.012,114.606)(3.349,109.557)
	(3.730,104.512)(4.156,99.471)(4.627,94.433)
	(5.142,89.400)(5.702,84.371)(6.307,79.348)
	(6.957,74.330)(7.650,69.319)(8.389,64.313)
	(9.171,59.315)(9.998,54.323)(10.870,49.339)
	(11.785,44.363)(12.745,39.395)(13.748,34.436)
	(14.796,29.487)(15.887,24.546)(17.023,19.616)
	(18.202,14.695)(19.424,9.786)(20.690,4.887)
	(22.000,0.000)
\drawline(322.000,-0.000)(323.310,4.887)(324.576,9.786)
	(325.798,14.695)(326.977,19.616)(328.113,24.546)
	(329.204,29.487)(330.252,34.436)(331.255,39.395)
	(332.215,44.363)(333.130,49.339)(334.002,54.323)
	(334.829,59.315)(335.611,64.313)(336.350,69.319)
	(337.043,74.330)(337.693,79.348)(338.298,84.371)
	(338.858,89.400)(339.373,94.433)(339.844,99.471)
	(340.270,104.512)(340.651,109.557)(340.988,114.606)
	(341.280,119.657)(341.527,124.710)(341.729,129.766)
	(341.886,134.823)(341.998,139.881)(342.065,144.940)
	(342.088,150.000)(342.065,155.060)(341.998,160.119)
	(341.886,165.177)(341.729,170.234)(341.527,175.290)
	(341.280,180.343)(340.988,185.394)(340.651,190.443)
	(340.270,195.488)(339.844,200.529)(339.373,205.567)
	(338.858,210.600)(338.298,215.629)(337.693,220.652)
	(337.043,225.670)(336.350,230.681)(335.611,235.687)
	(334.829,240.685)(334.002,245.677)(333.130,250.661)
	(332.215,255.637)(331.255,260.605)(330.252,265.564)
	(329.204,270.513)(328.113,275.454)(326.977,280.384)
	(325.798,285.305)(324.576,290.214)(323.310,295.113)
	(322.000,300.000)
\thinlines
\drawline(142,180)(142,280)(42,280)
	(42,180)(142,180)
\drawline(262,60)(262,120)(202,120)
	(202,60)(262,60)
\drawline(302,20)(302,280)(282,280)
	(282,20)(302,20)
\drawline(182,20)(182,280)(162,280)
	(162,20)(182,20)
\end{picture}
}
Notice that there are no rectangular boxes which 
span entire rows of the matrix $\Gamma$ above.

This intensity matrix $\Gamma$  gives quadratic relations 
which then have to be checked for
consistency for the $3$-site case.  We can do the same with 
another intensity matrix $\Gamma':=P_{\mu\nu}\Gamma P_{\mu\nu}$, 
where $P_{\mu\nu}$ is a permutation matrix,
$$P_{\mu\nu}(u_{\mu_k}\otimes u_{\nu_{\scriptstyle k+1}})
=(u_{\nu_k}\otimes u_{\mu_{\scriptstyle k+1}})$$
where $u_{\mu_k,\nu_k}$ are defined before eq.(1.15). 
The same procedure applies for any number of species $N$.

To sum up, we have shown that there are a number of quadratic 
algebras which allow the computation of the ground state wave-function.
What is missing is a complete classification of the algebras.
What is also missing is an understanding of the origin of the
degeneracies.  Are the ground state wave-functions some 
representations of some algebra?  In the special case where the 
Hamiltonian densities are generators of a Hecke algebra, the answer
is known, but not in the general case.  

We would also like to emphasize that the ground state wave-functions 
thus computed are non-trivial in the sense that connected 
correlations functions in these states are non-zero.  One simple example 
is already known. This is given in eq. (3.18) \cite{Schutz}.
The method used in \cite{Schutz} to 
compute the correlation functions was different.
Other cases can be considered using the results of this section.

\setcounter{equation}{0}
\section{Linear chain with left end open and right end closed}

As discussed in section 1, the ground state 
wave-function is given by the expression 
\eql(4.1)
\ket(P_s)=\langle 0| \prod_{k=1}^L \left( \sum_{\mu_k=0}^{N-1} 
D_{\mu_k}u_{\mu_k} \right),
\eq
where the $D_{\mu}$'s satisfy one of the polynomial algebras described in
the last section and a new condition (see eq. (1.22))
\eql(4.2)
\sum_{\beta=0}^{N-1}L_{\alpha}^{\beta}\langle 0| D_{\beta}=0,
\eq
where $L_{\alpha}^{\beta}$ is an intensity matrix with cofactors ${\cal
L}_{\beta}$.  We consider the cases $N=2$ and $N=3$ only.  The following
conditions on the $D$ matrices are obtained depending on the structure of
the boundary intensity matrix.  In each case, only the non-zero rates
are written down.

%\leftline{$\underline{N=2}$}
{$\underline{N=2}$}

\begin{enumerate}

\item
\eql(4.3)
\langle 0| D_1 = {L_1^0\over L_0^1} \langle 0| D_0\quad  
(L^1_0,L_1^0\neq 0) 
\eq

\item
\eql(4.4)
\langle 0| D_1 = 0 \qquad (L^1_0\neq 0) 
\eq

\item
\eql(4.5)
\langle 0| D_0 =0 \qquad (L_1^0\neq 0) 
\eq

\end{enumerate}

%\leftline{$\underline{N=3}$}
{$\underline{N=3}$}

For the case where at least one principal cofactor, e. g.  ${\cal L}_0$, 
 is non-zero 
\eql(4.6)
\langle 0| D_{\alpha} = {{\cal L}_{\alpha}\over {\cal L}_0} \langle 0| D_0
\;\; (\alpha=1,2)
\eq
and one obtains two constraints.  If all the principal cofactors are $0$,
only two rates are non-zero.  We can distinguish the following cases:

\begin{enumerate}

\item
\eql(4.7)
\langle 0| D_1 = {L_1^0\over
L_0^1} \langle 0 | D_0 \quad L_0^1, L_1^0 \neq 0; 
\eq

\item
\eql(4.8)
\langle 0| D_2 = {L_2^0\over
L_0^2} \langle 0 | D_0 \quad L_0^2, L_2^0 \neq 0; 
\eq

\item
\eql(4.9)
\langle 0| D_2 = {L_2^1\over
L_1^2} \langle 0 | D_1 \quad L_2^1,  L_1^2 \neq 0; 
\eq

\item
\eql(4.10)
\langle 0| D_1 = 0  \qquad (L_0^1+ L_2^1) \neq 0; 
\eq

\item
\eql(4.11)
\langle 0| D_0 = 0 \qquad (L_1^0 + L_2^0) \neq 0; 
\eq

\item
\eql(4.12)
\langle 0| D_2 = 0 \qquad (L_0^2 + L_1^2) \neq 0. 
\eq

\end{enumerate}

Notice that for the last six cases, one has only one constraint.  The
conditions (4.3) to (4.12) have to be compatible with the polynomial
algebras.  In this way one can, in some cases, find relations between the
bulk rates only.  In most cases the degeneracy of the ground state is
lifted. 

The marriage of the bulk algebra with the boundary conditions (4.3-12)
is a straightforward mathematical exercise. 
Therefore, we shall confine ourselves
to models of asymmetric diffusion where the only non-zero rates are
$\Ga(\alpha\beta,\beta\alpha)$ and give examples of how the concentration
profiles can be computed.  These are the cases for which the relevant
algebras are given in eq. (3.18) for $N=2$ and eq. (3.30) for $N=3$.  

We shall start with the $N=2$ case.  We have the algebra
\eql(4.13)
D_0 D_1 = q^2 D_1 D_0
\eq
and one of the conditions (4.3$-$5).  The conditions eq. (4.4) and 
(4.5) give a trivial ground state: $\langle 0|D_0^L$ (eq. 4.4) or
$\langle 0|D_1^L$ (eq. 4.5).  We shall rewrite eq. (4.3) as follows:
\eql(4.14)
\langle 0|D_1=\mu \langle 0|D_0.
\eq
There are no constraints connecting $q^2$ and $\mu$.  Assume that 
$D_1$ indicates the presence of a particle on any particular site
and $D_0$ the lack of one, {\em i.e.} holes.  
The concentration of particles at a distance $k$ from the left boundary
for a chain of length $L$ is \cite{DEHP}
\eql(4.15)
c(k)={1\over Z_L} \langle 0| C^{k-1}D_1 C^{L-k},
\eq
where 
\eql(4.16)
Z_L=\langle 0|C^L, \qquad C=D_0+D_1.
\eq
A straightforward calculation gives
\eql(4.17)
c(k)={\dis 1\over 1+{1\over \mu}q^{2(1-k)}}.
\eq
The connected two-point correlation function
\eql(4.18)
c(k,l)={1\over Z_L}\langle 0|\left( C^{k-1}D_1 C^{l-k-1} D_1 C^{L-l}\right)
 - c(k)c(l)
\eq
vanishes.

An interesting case occurs for the algebra
\eql(4.19)
D_1 D_0 =0
\eq
which corresponds to $\Ga(01,10)=0$ and consider the boundary condition
$L_1^0=0$ (eq. 4.4).  Upon inserting these conditions 
into eq. (4.15) we obtain
\eql(4.20)
c(k)={\langle 0|\sum_{j=1}^{k-1} D_0^{j} D_1^{L-j}\over
\langle 0|\sum_{j=1}^{L} D_0^j D_1^{L-j}}
\eq
and we cannot go any further because our algebraic rules do not 
allow us to evaluate this expression.  We leave it as an exercise
to the reader to find the physical reasons why the method appears not
to work in this case, and to do the calculation properly.

We now consider the  $N=3$ states problem and illustrate the 
method in the case of eq. (3.30), which we rewrite for convenience:
\eql(4.21)
q_0 D_1 D_2 = D_2 D_1. \quad q_1 D_2 D_0 = D_0 D_2, 
\quad q_2 D_0 D_1 = D_1 D_0,
\eq
where 
\eql(4.22)
q_0={\Ga(12,21)\over\Ga(21,12)},\quad q_1={\Ga(20,02)\over\Ga(02,20)},
\quad q_2={\Ga(01,10)\over\Ga(10,01)}.
\eq
We consider the boundary conditions (4.6) and assume all the rates
in (4.22) to be non-zero.  Setting 
$\mu_{\alpha}:={\cal L}_{\alpha}/{\cal L}_0$, we have
\eql(4.23)
\langle 0| D_1 = \mu_1 \langle 0| D_0 \;\;\;\mbox{and}\;\;\;
\langle 0| D_2 = \mu_2 \langle 0| D_0.
\eq

As we shall now show, consistency conditions on the rates show up. 
Let us first compute
\eql(4.24)
\begin{array}{ccccccc}
q_0\langle 0|D_1 D_2&=&q_0\mu_1\langle 0|D_0 D_2&=&
q_0 q_1 \mu_1\langle 0|D_2 D_0&=&q_0q_1\mu_1\mu_2\langle 0|D_0^2;\\
\langle 0|D_2 D_1&=&\mu_2\langle 0|D_0 D_1&=&
(\mu_2/q_2)\langle 0|D_1 D_0&=&(\mu_1\mu_2/q_2)\langle 0|D_0^2.
\end{array}
\eq
From the first equation in (4.21) and equations (4.24) we get
\eql(4.25)
q_0 q_1 q_2 = 1.
\eq
Note that this condition constrains only the bulk rates.  The 
concentration profiles for particles ``$1$'' and ``$2$'' (``$0$''
denotes vacancies) can easily be computed \cite{Evans-etal} as
$$c_1(k)={1\over Z_L}\langle 0| C^{k-1} D_1 C^{L-k};\qquad
c_2(k)={1\over Z_L}\langle 0| C^{k-1} D_2 C^{L-k}$$
\eql(4.26)
Z_L=\langle 0| C^L; \qquad C=D_0+D_1+D_2.
\eq
Using equations (4.21), (4.23) and (4.25) we obtain
\eql(4.27)
c_1(k)={\mu_1 q_2^{1-k}\over 1+\mu_1 q_2^{1-k} + \mu_2 q_1^{k-1}};
\quad {c_1(k)\over c_2(k)}={\mu_1\over \mu_2}q_0^{k-1}.
\eq
We are not aware of the existence of another method which would give 
us the expressions (4.27).  The connected two-point correlation
function is zero.

\setcounter{equation}{0}
\section{The open chain three-state exclusion models}

This is a much more difficult problem.
We limit our investigation to exclusion models in which the only
non-vanishing rates in the bulk are
\eql(5.1)
g_{\alpha\beta}:=\Ga(\alpha\beta,\beta\alpha)
\eq
and make the assumption that the matrices $X_{\mu}$'s in equation (1.12)
are c-numbers, which we shall call $x_{\mu}$.  
As we shall see below, the choice $x_{\mu}=0$ is possible for 
pathological choices of boundary conditions only.
The choice (5.1) is not only
justified by the interest in these physical processes but as shown later,
(see the discussion after eq. (5.24)) in this way 
one obtains quadratic algebras
which are understood \cite{Zvyagina,Vershik}.

We are looking therefore at
Fock representations of the algebra
\eql(5.2)
\begin{array}{rcl}
g_{01}D_0 D_1 - g_{10}D_1 D_0 &=& x_0 D_1 - x_1 D_0 \\
g_{12}D_1 D_2 - g_{21}D_2 D_1 &=& x_1 D_2 - x_2 D_1 \\
g_{20}D_2 D_0 - g_{02}D_0 D_2 &=& x_2 D_0 - x_0 D_2 
\end{array}
\eq
with boundary conditions on the states $\langle 0|$ and $|0\rangle$
\eql(5.3)
\langle 0| (x_{\nu}-L_{\nu}^{\mu}D_{\mu})=0 \quad \mbox{and}\quad
(x_{\nu}+R_{\nu}^{\mu}D_{\mu})|0\rangle=0.
\eq
As shown in ref. \cite{EssRit}, for $N=2$, one has Fock representations of
the algebra for any bulk and boundary rates.  The situation is going to be
very different for $N=3$.  The reason is that the number of boundary
conditions determined by the matrices $L_{\mu}^{\nu}$ and $R_{\mu}^{\nu}$
is in general too large.  For each of $L$ and $R$ one has to consider
all the cases enumerated in the previous section (see eqs. (4.6$-$12)).
We now make some transformations which exploit the fact that the 
$x_{\mu}$'s are still free parameters.

Let us first consider words of length $1$:
\eql(5.4)
\langle 0| D_{\mu}|0\rangle=\delta_{\mu}.
\eq
From equation (5.3), it follows that
\eql(5.5)
(L^{\mu}_{\nu}+R^{\mu}_{\nu})\delta_{\mu}=0
\eq
and
\eql(5.6)
x_{\nu}={1\over 2}(L^{\mu}_{\nu}-R^{\mu}_{\nu})\delta_{\mu}.
\eq
We have to keep in mind that 
\eql(5.7)
B_{\mu}^{\nu}=L_{\mu}^{\nu}+R_{\mu}^{\nu}
\eq
is an intensity matrix with cofactors ${\cal B}_{\mu}$.  This implies (using
eqs. (2.3-6)) that
\eql(5.8)
\delta_{\mu}=\xi{\cal B}_{\mu}
\eq
and
\eql(5.9)
\begin{array}{rcl}
x_{\nu}&=&\dis {\xi\over 2}(L^{\mu}_{\nu}-R^{\mu}_{\nu}){\cal B}_{\mu}\\
&=&\xi(L_{\nu}^{\mu}{\cal R}_{\mu}-R_{\nu}^{\mu}{\cal L}_{\mu}).
\end{array}
\eq
Here $\xi$ is an arbitrary parameter and ${\cal R}_{\mu}({\cal L}_{\mu})$
are the cofactors of the intensity matrix $R_{\nu}^{\mu}(L_{\nu}^{\mu})$.
The expressions for the various cofactors can be obtained using eq. (2.7).
Since $R_{\nu}^{\mu}$ and $L_{\nu}^{\mu}$ are intensity matrices, it
follows from eq.(5.3) that
\eql(5.10)
\sum_{\nu=0}^2x_{\nu}=0.
\eq

It is interesting to note that if we have a Fock representation, the ratio
of the currents of particles $1$ and $2$ is already known.  It is easy to 
show that 
\eql(5.11)
J_i=x_i{\langle0|C^{L-1}|0\rangle\over\langle 0|C^{L}|0\rangle}, \;\; i=1,2
\eq
where
\eql(5.12)
C=\sum_{\mu=0}^2 D_{\mu}.
\eq
This implies
\eql(5.13)
{J_1\over J_2}={x_1\over x_2}
\eq
where $x_1/x_2$ is given by the boundary conditions (see eq. (5.9)).

In deriving (5.11) we have used the following expression for the current
density \cite{Evans-etal,DEHP} 
\eql(5.14)
\begin{array}{rcl}
j_1&=&g_{10}D_1 D_0 + g_{12} D_1 D_2 - g_{01} D_0 D_1 - g_{21} D_2 D_1\\
&=&-(x_0+x_2)D_1 + x_1 (D_0+D_2)=x_1 C,
\end{array}
\eq
and a similar equation for $j_2$.  We also used the definition
\eql(5.15)
J_i=\langle j_i\rangle.
\eq

It is easy to find the one-dimensional representations of the algebra (5.2).
Using eq. (5.4), we get the following conditions on the rates:
\eql(5.16)
\begin{array}{ccc}
g_{01}-g_{10}&=&y_0-y_1\\
g_{12}-g_{21}&=&y_1-y_2\\
g_{20}-g_{02}&=&y_2-y_0
\end{array}
\eq
where 
\eql(5.17)
y_{\mu}={x_{\mu}\over \delta_{\mu}}
\eq
are given by the boundary conditions (see eqs. (5.8) and (5.9)).  Notice
that eqs. (5.16) give one relation between the bulk rates:
\eql(5.18)
g_{01}+g_{12}+g_{20}=g_{10}+g_{21}+g_{02}
\eq
and two relations between boundary and bulk rates.

Before we start looking for the cases in which the Fock representations of
the algebra (5.2) exist, it is useful to bring it into a different form.  We
denote
\eql(5.19)
u_0={y_1-y_2\over g_{21}}+1-q_0, \quad u_1={y_2-y_0\over g_{02}}+1-q_1,
\quad u_2={y_0-y_1\over g_{10}}+1-q_2
\eq
and also
\eql(5.20)
\begin{array}{cc}
\dis v_{10}=1-q_2+(y_0/g_{10})&\dis v_{01}=1-q_2-(y_1/g_{10})\\
\dis v_{21}=1-q_0+(y_1/g_{21})&\dis v_{12}=1-q_0-(y_2/g_{21})\\
\dis v_{02}=1-q_1+(y_2/g_{02})&\dis v_{20}=1-q_1-(y_0/g_{02}).
\end{array}
\eq
The $q_{\mu}$'s and $y_{\nu}$'s in the above are defined in equations (4.22)
and (5.17) respectively.  We also define new generators $H_{\mu}$
\eql(5.21)
D_{\mu}=\delta_{\mu}(1+H_{\mu}).
\eq

With these new definitions, the algebra relations and the action of the
generators on the vacuum are
\eql(5.22)
\begin{array}{ccc}
q_2 H_0 H_1 - H_1 H_0 &=& u_2 + v_{10}H_1 + v_{01}H_0\\
q_0 H_1 H_2 - H_2 H_1 &=& u_0 + v_{21}H_2 + v_{12}H_1\\
q_1 H_2 H_0 - H_0 H_2 &=& u_1 + v_{02}H_0 + v_{20}H_2
\end{array}
\eq
with
\eql(5.23)
\langle 0| H_0 |0\rangle = \langle 0| H_1 |0\rangle = 
\langle 0| H_2 |0\rangle = 0,
\eq
\eql(5.24)
\begin{array}{cc}
({\cal L}_0/\delta_0)\langle 0| H_1 = ({\cal L}_1/\delta_1)\langle 0|H_0;&
({\cal L}_0/\delta_0)\langle 0| H_2 = ({\cal L}_2/\delta_2)\langle 0|H_0\\
({\cal R}_0/\delta_0)H_1|0\rangle   = ({\cal R}_1/\delta_1)H_0|0\rangle;&
({\cal R}_0/\delta_0)H_2|0\rangle   = ({\cal R}_2/\delta_2)H_0|0\rangle.
\end{array}
\eq

The algebra (5.22) is of a special case of the ones defined in
eq. (1.24) \cite{Zvyagina,Vershik}.  For the latter one can show that
For the latter one can show that if the coefficients are
generic, then for $N\geq 3$ the algebra is finite dimensional.  For example
there are $28$ independent monomials for $N=3$.  Therefore one can already
understand that the $N=3$ problem is different from the one for $N=2$.  The
next point to note is the role of boundary rates for $N=3$.  They appear
explicitly in eq. (5.24) but also through the $y_{\mu}$'s in the
coefficients of the algebra (5.22) (see eq. (5.20)).

In the next section we ask the question if the algebra (5.22) has
representations in the generic case (when the principal cofactors of the
boundary matrices are non-zero) aside from the one-dimensional one, which
corresponds to $u_{\mu}=0$ and $H_{\mu}=0$.  We shall show that 
one can have representations of dimension at most $2$.  In general, it 
is not possible to have higher dimensional representations because 
the number of constraints coming from the algebra relations is larger that
the number of boundary and bulk rates.  In the subsequent
sections, we will make a systematic
investigation of the cases where some or all of the principal minors 
vanish.   

\setcounter{equation}{0}
\section{Generic 3-state exclusion model: boundary intensity 
matrices have non-zero minors}

In this section we consider the question when the algebra (5.22) 
has Fock representations 
defined by eqs (5.23$-$4) under the assumption that the ${\cal R}_{\mu}$
and ${\cal L}_{\mu}$ are non-zero.  This is a very technical 
section and the reader not interested in the method of answering
the question can proceed directly to the end of the section
where the result is given.  For our specific purpose it is 
convenient to rewrite the algebra in a slightly different way 
using new notation.

Let 
\eql(6.1)
q_{\nu}=Q_{\nu}^2;\;\;\lambda_i={{\cal L}_i\over {\cal L}_0}
{\delta_0\over \delta_i},\;\;\mu_i={{\cal R}_i\over {\cal R}_0}
{\delta_0\over \delta_i},\;\;\alpha_i^2={\lambda_i\over\mu_i},
\quad(\nu=0,1,2;\;i=1,2)
\eq
and instead of the generators $H_{\nu}$ use $L_{\nu}$ defined
as follows:
\eql(6.2)
D_0=\delta_0(1+L_0),\quad D_i=\delta_1(1+\lambda_1 L_1),\quad
D_2=\delta_2(1+\mu_2 L_2).
\eq

The algebra relations are
\eql(6.3)
\begin{array}{rcl}
Q^2_2 L_0 L_1 - L_1 L_0 &=& m_2 + n_{10} L_1 + n_{01} L_0 \\
Q^2_0 L_1 L_2 - L_2 L_1 &=& m_0 + n_{21} L_2 + n_{12} L_1 \\
Q^2_1 L_2 L_0 - L_0 L_2 &=& m_1 + n_{02} L_0 + n_{20} L_2 
\end{array}
\eq
and the conditions for a Fock representation are
\eql(6.4)
\begin{array}{cccccc}
\langle 0|L_1&=&\langle 0|L_0,&\langle 0|L_2&=&\alpha_1^2 \langle 0|L_0,\\
L_2|0\rangle&=&L_0|0\rangle,&L_1|0\rangle&=&\alpha_2^2 L_0|0\rangle,\\
\langle 0| L_{\nu}|0\rangle&=&0,& \;\;(\nu=0,1,2)&&.
\end{array}
\eq
This algebra depends on $14$ parameters, and after setting
the time scale, we are effectively left with $13$.  For the
physical problem, the $m_{\nu}$'s and $n_{\mu\nu}$'s are
related to the rates:
\eql(6.5)
\begin{array}{rcl}
m_0&=&\dis {1\over \lambda_1\mu_2}\left({y_1-y_2\over g_{12}}+1-Q_0^2\right),\\
m_1&=&\dis {1\over \mu_2}\left({y_2-y_0\over g_{02}}+1-Q_1^2\right),\\
m_2&=&\dis {1\over \lambda_1}\left({y_0-y_1\over g_{10}}+1-Q_2^2\right)
\end{array}
\eq
and
\eql(6.6)
\begin{array}{rclrcl}
n_{10}&=&\dis 1-Q_2^2+{y_0\over g_{10}},
&n_{01}&=&\dis {1\over\lambda_1}\left(1-Q_2^2-{y_1\over g_{10}}\right),\\
n_{02}&=&\dis {1\over\mu_2}\left(1-Q_1^2+{y_2\over g_{02}}\right),&
n_{20}&=&\dis \left(1-Q_1^2-{y_0\over g_{02}}\right),\\
n_{21}&=&\dis {1\over\lambda_1}\left(1-Q_0^2+{y_1\over g_{21}}\right),&
n_{12}&=&\dis {1\over\mu_2}\left(1-Q_0^2-{y_2\over g_{21}}\right).
\end{array}
\eq

At this point two strategies are possible.  One can look for 
vacuum expectation values of words of different lengths and find 
consistency conditions for the $14$ paramenters; or, one can look 
for matrix representations of the algebra and in this way
obtain consistency conditions among the parameters.  We shall do 
both -- we shall take matrix representations and indicate which of the
consistency conditions on the matrix elements also come
from specific vacuum expectation values of words of given length.
In order to do so, it is convenient to work with paramaters $p_{\mu\nu}$
and $r_{\mu}$ introduced below instead of the $m_{\mu}$ and $n_{\mu\nu}$
introduced earlier.
\eql(6.7)
\begin{array}{ccl}
m_0&=&Q_0\alpha_1\alpha_2 N(Q_0/(\alpha_1\alpha_2))r_0,\\
m_1&=&Q_1\alpha_1 N(Q_1\alpha_1)r_1,\\
m_2&=&Q_2\alpha_2 N(Q_2\alpha_2)r_2,\\
n_{01}&=&Q_2\alpha_2 N(Q_2^2\alpha_2)p_{01},\\
n_{10}&=&Q_2 N(Q_2^2\alpha_2)p_{10},\\
n_{20}&=&Q_1 N(Q_1^2\alpha_1)p_{20},\\
n_{02}&=&Q_1\alpha_1 N(Q_1^2\alpha_1)p_{02},\\
n_{21}&=&Q_0\alpha_2 N(Q_0^2/(\alpha_1\alpha_2))p_{21},\\
n_{12}&=&Q_0\alpha_1 N(Q_0^2/(\alpha_1\alpha_2))p_{12}.
\end{array}
\eq
The constraints on the parameters we shall obtain in what follows
look much simpler in terms of the new parameters.  In eqs. (6.7)
and in the what follows we use 
the notation
\eql(6.8)
M(x)=x+{1\over x}\quad\mbox{and}\quad N(x)=x-{1\over x}.
\eq
We consider a three-dimensional representation of the algebra
eq. (6.3).  We make this choice since we are going to prove
that in the the generic case, the two-dimensional representation 
is the largest we can have.
Using the boundary conditions (6.4) we can bring
the matrix representation of the generators $L_{\mu}$ to a 
tridiagonal form by means of a similarity transformation
by an orthogonal matrix, just as it was done for the $N=2$ case in 
\cite{EssRit}.  One can show that with 
$$\langle 0|\stackrel{\cdot}{=} (1,0,0)\quad \hbox{and}\quad
|0\rangle \stackrel{\cdot}{=} (1,0,0)^T$$
(where $T$ denotes transposition) the generators then take on the 
expressions
$$L_0\stackrel{\cdot}{=}\left(
\begin{array}{ccc}
0&\sqrt{f_1}&0\\
\sqrt{f_1}&c_1^{(0)}&\sqrt{f_2}\\
0&\sqrt{f_2}&c_2^{(0)}
\end{array}\right)$$
\eql(6.9)
L_1\stackrel{\cdot}{=}\left(
\begin{array}{ccc}
0&\sqrt{f_1}&0\\
\alpha_2^2\sqrt{f_1}&\alpha_2 c_1^{(1)}&\dis {\sqrt{f_2}\over Q_2^2}\\
0&\alpha_2^2 Q_2^2\sqrt{f_2}&\alpha_2c_2^{(1)}
\end{array}\right)
\eq
$$L_2\stackrel{\cdot}{=}\left(
\begin{array}{ccc}
0&\alpha_1^2\sqrt{f_1}&0\\
\sqrt{f_1}&\alpha_1c_1^{(2)}&\alpha_1^2 Q_1^2\sqrt{f_2}\\
0&\dis {\sqrt{f_2}\over Q_1^2}&\alpha_1 c_2^{(2)}
\end{array}\right)$$

Notice that a one-dimensional representation is obtained if $f_1=0$,
a two-dimensional one if $f_2=0,f_1\neq 0$ and a three-dimensional
one if $f_1\neq 0,f_2\neq 0$.  $f_i$ and $c^{(j)}_i$ are still to be 
determined from the algebra relations, the Fock conditions (6.4) 
having already been taken into account.  

We now insert the 
$3\times 3$ matrices (6.9) in the algebraic relations (6.3) and
from the equality of each matrix element $(i,j)$ on the left- and
right-hand sides, get the following equations.  From the $(1,1)$
matrix element, we obtain
\eql(6.10)
r_0=r_1=r_2=f_1.
\eq
From the matrix elements $(1,2)$, $(2,1)$ and $(2,2)$ we get
\eql(6.11)
\begin{array}{rclcl}
c_1^{(0)}&=&M(\alpha_2 Q_2)p_{10} + M(Q_2) p_{01}&=&
M(Q_1)p_{02}+M(\alpha_1Q_1)p_{20}\\
c_1^{(1)}&=&M(Q_2)p_{10} + M(\alpha_2 Q_2) p_{01}&=&
M(Q_0/(\alpha_1\alpha_2))p_{21}+M(Q_0)p_{12}\\
c_1^{(2)}&=&M(\alpha_1Q_1)p_{02} + M(Q_1) p_{20}&=&
M(Q_0/(\alpha_1\alpha_2))p_{12}+M(Q_0)p_{21}\\
\end{array}
\eq
\eql(6.12)
\begin{array}{lcl}
\dis {N(Q_2^3\alpha_2)\over N(\alpha_2)M(Q_2)}f_2&=&
f_1+p_{10}^2+p_{01}^2+M(Q_2^2\alpha_2)p_{01}p_{10}\\
\dis {N((\alpha_1\alpha_2)/Q_0^3)\over 
N(\alpha_1\alpha_2)M(Q_0)}f_2&=&
f_1+p_{12}^2+p_{21}^2+M((\alpha_1\alpha_2)/Q_0^2)p_{21}p_{12}\\
\dis {N(Q_1^3\alpha_1)\over N(\alpha_1)M(Q_1)}f_2&=&
f_1+p_{20}^2+p_{02}^2+M(Q_1^2\alpha_1)p_{02}p_{20}.
\end{array}
\eq

Let us pause here for a moment.  Equations (6.10) and (6.11)
can also be obtained from the vacuum expectation values of 
words of length two and three.  The $1$-dimensional representation
is obtained from the condition $f_1=0$ and the $2$-dimensional 
representation is obtained by setting $f_2=0$ in (6.12). From 
eqs. (6.11) and (6.12) the $p_{\mu\nu}$ are completely
determined as are the $r_{\mu}$ (eq. (6.10)).  This means that in the
algebra (6.3), the free parameters are $\alpha_1, \alpha_2$ and
$Q_{\mu},\mu=0,1,2$.  The next logical step would be to solve the
equations and look for solutions with positive rates.  This is a 
difficult exercise.  For a given physical problem, where some 
conditions on the rates are given, the problem is simpler since there 
are fewer parameters.

We now go on with the rest of the matrix elements $(1,3)$, $(2,3)$,
$(3,1)$, $(3,2)$ and $(3,3)$.  We get:
\eql(6.13)
Q_0^2 Q_1^2 Q_2^2 = 1,
\eq
\eql(6.14)
\begin{array}{rcl}
c_2^{(0)}&=&\dis {M(Q_1)\over N(\alpha_1 Q_1^4)}\Bigl(
\bigl(N(\alpha_1^2 Q_1^4) - N(Q_1^2)\bigr)p_{20} + 
M(Q_1)N(Q_1\alpha_1)p_{02}\Bigr)\\
&=&\dis {M(Q_2)\over N(\alpha_2 Q_2^4)}\Bigl(
\bigl(N(\alpha_2^2 Q_2^4) - N(Q_2^2)\bigr)p_{10} + 
M(Q_2)N(Q_2\alpha_2)p_{01}\Bigr)\\
c_2^{(1)}&=&\dis {M(Q_2)\over N(\alpha_2 Q_2^4)}\Bigl(
\bigl(N(\alpha_2^2 Q_2^4) - N(Q_2^2)\bigr)p_{01} + 
M(Q_2)N(Q_2\alpha_2)p_{10}\Bigr)\\
&=&\dis {M(Q_0)\over N(Q_0^4/(\alpha_1\alpha_2))}\Bigl(
\bigl(N({Q^4_0\over\alpha_1\alpha_2}) - N(Q_0^2)\bigr)p_{21} + 
M(Q_0)N({Q_0\over\alpha_1\alpha_2})p_{12}\Bigr)\\
c_2^{(2)}&=&\dis {M(Q_1)\over N(\alpha_1 Q_1^4)}\Bigl(
\bigl(N(\alpha_1^2 Q_1^4) - N(Q_1^2)\bigr)p_{02} + 
M(Q_1)N(Q_2\alpha_2)p_{20}\Bigr)\\
&=&\dis {M(Q_0)\over N({Q_0^4\over\alpha_1\alpha_2})}\Bigl(
\bigl(N({Q^4_0\over\alpha_1\alpha_2}) - N(Q_0^2)\bigr)p_{12} + 
M(Q_0)N({Q_0\over\alpha_1\alpha_2})p_{21}\Bigr)
\end{array}
\eq
and
\eql(6.15)
\begin{array}{rcl}
f_2&=&\dis -f_1+{1\over N(\alpha_2 Q_2)}\Bigl(
N(Q_2)c_0^{(2)}c_1^{(2)}-N(\alpha_2 Q_2^2)
(p_{01}c_0^{(2)}+p_{10}c_1^{(2)})
\Bigr)\\
&=&\dis -f_1+{1\over N({Q_0\over\alpha_1\alpha_2})}\Bigl(
N(Q_0)c_1^{(2)}c_2^{(2)}-N(Q_0^2/(\alpha_1\alpha_2))
(p_{12}c_1^{(2)}+p_{21}c_2^{(2)})
\Bigr)\\
&=&\dis -f_1+{1\over N(\alpha_1 Q_1)}\Bigl(
N(Q_1)c_0^{(2)}c_2^{(2)}-N(\alpha_1 Q_1^2)
(p_{20}c_2^{(2)}+p_{02}c_0^{(2)})
\Bigr).
\end{array}
\eq
Equations (6.11) and (6.14) give a system of six homogeneous equations
for the $p_{\mu\nu}$ with determinant generically different from zero.
In order to have a non-trivial solution for the $p_{\mu\nu}$'s, we
must set this determinant to zero, thus introducing an extra condition
on the rates.  This leaves one of them (say $p_{10}$) free.  
The same equations 
determine $c_i^{(\mu)}$ ($i=1,2$, $\mu=0,1,2$).  We are thus left with 
eight equations: three from (6.12), one from (6.13), three from (6.15),
and one determinantal condition 
for the seven unknowns $f_2,p_{10},Q_{\mu}$ and $\alpha_i$.  (We can
take $f_1=1$ in eq. (6.10) to set the time scale.)  This implies that
we cannot, generically have a $3$-dimensional representation.
This also implies that the algebra (6.3) with the Fock conditions
(6.4) can have representations of dimensions $1$ and $2$ only.  
The conditions on the parameters are not neat but this was so 
even for the case $N=2$
\cite{EssRit}.  

We now come back to the physical problem.  The algebra (5.22) contains
$9$ parameters $-$ $q_0,q_1,q_2$ and the ratios $(y_0/g_{10})$, 
$(y_0/g_{02})$, $(y_1/g_{10})$, $(y_1/g_{21})$, $(y_2/g_{02})$ and 
$(y_2/g_{21})$; the boundary conditions (5.24) depend on $4$ others,
the $\lambda_i$'s and $\mu_i$'s for $i=1,2$.  This makes a total of
$13$ parameters.  Therefore for the physical problem, 
the case of $2$-dimensional
representations stays interesting since the algebraic varieties on which 
the solution set exists is not trivial.  

\setcounter{equation}{0}
\section{The two non-zero cofactors case}

In the last section we considered the generic case where all the cofactors
${\cal L}_{\mu}$ and ${\cal R}_{\mu}$ are non-zero and we found that the
algebra (5.22) had Fock representations defined by eqs. (5.23) and (5.24) of
dimension $1$ and $2$ only.  We should have continued our study and
looked at all cases when some or all of the cofactors vanish.  In this
section we shall confine ourselves only to the case
\eql(7.1)
{\cal L}_0={\cal L}_2={\cal R}_0={\cal R}_1=0, 
\quad {\cal L}_1\neq 0, {\cal R}_2\neq 0.
\eq
We had in mind the physical application of ref. \cite{Evans-etal}.  The
interested reader can easily duplicate the calculations presented in
this section for the other cases.  With (7.1), the Fock representations have
to satisfy the following simple simple conditions
\eql(7.2)
\langle 0| H_0=\langle 0| H_2 = 0; \quad H_0|0\rangle=H_1|0\rangle=0, 
\quad \langle 0 | H_{\mu} |0\rangle=0.
\eq

In order to find out in which cases the algebra (5.22) has Fock
representations defined by eq. (7.2), we take vacuum expectation values for
monomials of different degree.  

\noindent {} Monomials of degree $2$ ({} consider the algebra
relations between $\langle 0|$ and $|0\rangle$) give:
\eql(7.3)
u_1=u_2=0.
\eq

\noindent {} Monomials of degree $3$ (consider $\langle 0|
 H_1 H_0 H_2 |0\rangle$ and commute $H_0$ through to the left and 
the right) give
\eql(7.4)
v_{10}=v_{20}=v.
\eq
\noindent {} Monomials of degree $4$, $\langle 0| H_1 H_1 H_0 H_2|0\rangle$ 
and $\langle 0| H_1 H_2 H_0 H_2|0\rangle$
give
\eql(7.5)
q_0 v_{01} = q_2 v_{21}\quad\mbox{and}\quad q_0 v_{02} = q_1 v_{12}
\eq
respectively.
We have to distinguish several cases

\noindent $\bullet$ a) $v_{12}$ and $v_{21}$ are non-zero.  

\noindent {} Monomials of degree $5$ (consider 
$\langle 0| H_1 H_0 H_1 H_2 H_2 |0\rangle$) give
\eql(7.6)
q_1=q_2=q.
\eq

Taking into account relations (7.3$-$6) the algebra (5.22) becomes
\eql(7.7)
\begin{array}{rcl}
\dis
q H_0 H_1 - H_1 H_0 &=&\dis v H_1 + {q\over q_0}v_{21} H_0\\
q H_2 H_0 - H_0 H_2 &=& \dis 
{q\over q_0}v_{12} H_0 + v H_2\\
q_0 H_1 H_2 - H_2 H_1 &=& u_0 + v_{12} H_1 +v_{21} H_2
\end{array}
\eq
It is useful to change variables
\eql(7.8)
H_0=-v{\cal N},\;\; H_1={u_0\over v_{12}}{\cal A},\;\;
H_2={u_0\over v_{21}}{\cal B},\;\;z_2={u_0 q_0\over v_{12}v_{21}}
\eq
in terms of which, the algebra becomes
\eql(7.9)
\begin{array}{lccc}
a)\qquad&\dis z_2\biggl({\cal A}{\cal B}-{1\over q_0} 
{\cal B}{\cal A}\biggr)
&=&1+{\cal A}+{\cal B}\\
b)\qquad&\dis {\cal A}{\cal N}-q {\cal N}{\cal A}&=&\dis 
{\cal A}-{\dis q\over z_2}{\cal N}\\
c)\qquad&{\cal N}{\cal B}-q {\cal B}{\cal N}&=&
\dis {\cal B}-{\dis q\over z_2}{\cal N}.
\end{array}
\eq
This algebra has the representations
\eql(7.10)
\langle 0|{\cal N}=0={\cal N}|0\rangle,\quad 
\langle 0|{\cal B}=0={\cal A}|0\rangle.
\eq

We can now use the results of ref. \cite{EssRit} to find the following
representation of (7.9a):
\eql(7.11)
{\cal A}\stackrel{\cdot}{=}\left(\begin{array}{cccc}
a_1&f_1&0&0\ldots\\
0&a_2&f_2&0\ldots\\
0&0&a_3&f_3\ldots\\
0&0&0&a_4\ldots\\
\vdots&\vdots&\vdots&\ddots
\end{array}\right),\quad
{\cal B}={\cal A}^T\stackrel{\cdot}{=}\left(\begin{array}{cccc}
a_1&0&0&0\ldots\\
f_1&a_2&0&0\ldots\\
0&f_2&a_3&0\ldots\\
0&0&f_3&a_4\ldots\\
\vdots&\vdots&\vdots&\ddots
\end{array}\right)
\eq
where
\eql(7.12)
\begin{array}{cc}
a_n = z_2^{-1}\{n-1\}_{\lambda},&f_n^2 = 
z_2^{-2}\{n\}_{\lambda}(z_2+\{n-1\}_{\lambda}),\\ 
\{n\}_{\lambda} = \dis
{1-\lambda^n\over 1-\lambda}, \; n\geq 1, \,\{0\}_{\lambda}=1&
\lambda = q_0^{-1}.
\end{array}
\eq

We have introduced the symbol $z_2$ in keeping with the notation of ref.
\cite{EssRit}.  ${\cal A}^T$ is the transpose of matrix ${\cal A}$.  There
are two cases when we can have representations for ${\cal N}$ in eqs.
(7.9):
\eql(7.13)
a.1)\qquad q=q_0,\quad v_{01}=v_{21},\quad v_{02}=v_{12}
\eq
when 
\eql(7.14)
{\cal N}\stackrel{\cdot}{=}\left(\begin{array}{cccc}
h_1&0&0&\ldots\\
0&h_2&0&\ldots\\
0&0&h_3&\ldots\\
\vdots&\vdots&\vdots&\ddots
\end{array}\right)
\eq
with 
\eql(7.15)
h_n=\{n-1\}_q
\eq
and
\eql(7.16)
a.2)\qquad q=0, \; \; q_0\neq 0,\quad v_{01}=v_{02}=0
\eq
with
\eql(7.17)
h_n=\{n-1\}_{q\rightarrow 0}=1-\delta_{n 1}
\eq

\noindent $\bullet$ b) We consider the case (see eq. (7.5))
\eql(7.18)
v_{01}=v_{02}=v_{12}=v_{21}=0,\quad v_{10}=v_{20}=v,\quad u_1=u_2=0.
\eq

\noindent {} Monomials of degree $5$ give (see eq. (7.6))
\eql(7.19)
q_1=q_2=q.
\eq
With (7.18) and (7.19) the algebra (5.22) becomes 
\eql(7.20)
\begin{array}{ccc}
q H_0 H_1 - H_1 H_0 &=& v H_1\\
q H_2 H_0 - H_0 H_2 &=& v H_2\\
q_0 H_1 H_2 - H_2 H_1 &=& u_0.
\end{array}
\eq
We make the transformation 
\eql(7.21)
H_0=-v{\cal N},\;\; H_1=\sqrt{{u_0\over q_0}}{\cal A},\;\;
H_2=\sqrt{{u_0\over q_0}}{\cal B},\;\;\lambda={1\over q_0}
\eq
and get the algebra relations
\eql(7.22)
\begin{array}{ccc}
{\cal A}{\cal B}-\lambda {\cal B}{\cal A}&=&1\\
{\cal A}{\cal N}-q {\cal N}{\cal A}&=&{\cal A}\\
{\cal N}{\cal B}-q {\cal B}{\cal N}&=&{\cal B}
\end{array}
\eq
and
\eql(7.23)
{\cal A}|0\rangle={\cal N}|0\rangle = 0 =
\langle 0|{\cal B}=\langle 0|{\cal N}.
\eq
The Fock representation of this algebra (matrices satisfying (7.22) and
(7.23)) is:
\eql(7.24)
{\cal A}={\cal B}^T\stackrel{\cdot}{=}\left(\begin{array}{cccc}
0&g_1&0&\ldots\\
0&0&g_2&\\
0&0&0&\\
\vdots&&&\ddots
\end{array}\right),\quad
{\cal N}\stackrel{\cdot}{=}\left(\begin{array}{cccc}
h_1&0&0&\ldots\\
0&h_2&0&\\
0&0&h_3&\\
\vdots&&&\ddots
\end{array}\right)
\eq
where
\eql(7.25)
g_n^2=\{n\}_{\lambda},\quad h_n=\{n-1\}_q.
\eq

\noindent $\bullet$ c) We consider the case (see (eq.(7.5))
$$v_{21}=v_{01}=0.$$
The algebra (5.22) again with $q_1=q_2=q$ becomes
\eql(7.26)
\begin{array}{rcl}
q H_0 H_1 - H_1 H_0 &=& v H_1\\
q H_2 H_0 - H_0 H_2 &=&\dis {q\over q_0}v_{12} H_0 + v H_2\\
q_0 H_1 H_2 - H_2 H_1 &=& u_0 + v_{12} H_1 
\end{array}
\eq
We make the substitutions
\eql(7.27)
H_0=-v{\cal N},\;\; H_1={u_0\over v_{12}}{\cal A},\;\;
H_2={v_{12}\over q_0}{\cal B},\;\;\lambda={1\over q_0}
\eq
to get the algebra
\eql(7.28)
\begin{array}{ccc}
{\cal A}{\cal B}-\lambda {\cal B}{\cal A}&=&1+{\cal A}\\
{\cal A}{\cal N}-q {\cal N}{\cal A}&=&{\cal A}\\
{\cal N}{\cal B}-q {\cal B}{\cal N}&=&{\cal B}-q{\cal N}.
\end{array}
\eq
This algebra has the representations
\eql(7.29)
{\cal A}\stackrel{\cdot}{=}\left(\begin{array}{cccc}
0&f_1&0&\ldots\\
0&0&f_2&\\
0&0&0&\\
\vdots&&&\ddots
\end{array}\right),\quad
{\cal B}\stackrel{\cdot}{=}\left(\begin{array}{cccc}
b_1&0&0&\ldots\\
f_1&b_2&0&\\
0&f_2&b_3&\\
\vdots&&&\ddots
\end{array}\right)
\eq
where
\eql(7.30)
f_n^2=\{n\}_{\lambda},\quad b_n=\{n-1\}_{\lambda}
\eq
and if 
$$c.1)\qquad q=q_0=\lambda^{-1},$$
then
\eql(7.31)
{\cal N}\stackrel{\cdot}{=}\left(\begin{array}{cccc}
h_1&0&0&\ldots\\
0&h_2&0&\\
0&0&h_3&\\
\vdots&&&\ddots
\end{array}\right)
\eq
where
\eql(7.32)
h_k=\{k-1\}_{\lambda}.
\eq
If 
$$c.2) \qquad  q=0, $$ 
which implies $v_{02}=0$ one has to take 
\eql(7.33)
h_k=\{k-1\}_{\lambda\rightarrow 0}=1-\delta_{k 1}.
\eq

In order to help the reader, we shall summarize the results.  The algebra
(5.22) has Fock representations defined by
\eql(7.34)
\langle 0| H_2=\langle 0| H_0=0=H_0|0\rangle=H_1|0\rangle
\eq
if ({\em necessary condition})
\eql(7.35)
u_1=u_2=0,\quad v_{10}=v_{20}=v,\quad q_1=q_2=q,
\eq
in which case the algebra relations become
\eql(7.36)
\begin{array}{rcl}
q H_0 H_1 - H_1 H_0 &=& v H_1 + v_{01} H_0\\
q H_2 H_0 - H_0 H_2 &=& v_{02} H_0 + v H_2\\
q_0 H_1 H_2 - H_2 H_1 &=& u_0 + v_{12} H_1 +v_{21} H_2
\end{array}
\eq
This algebra has Fock representations in five cases

\begin{enumerate}

\item 

\eql(7.37)
v_{01}=v_{21},\quad v_{02}=v_{12},\quad q_0=q.
\eq
With the notations
$$H_0=-v{\cal N},\;\; H_1={u_0\over v_{12}}{\cal A},\;\;
H_2={u_0\over v_{21}}{\cal B},\;\;z_2={u_0 q_0\over v_{12}v_{21}},$$
${\cal A}$ and ${\cal B}$ are given by eq. (7.11-12) with $\lambda=q_0^{-1}$
and ${\cal N}$ by eqs. (7.14) with (7.15).

\item 

\eql(7.38)
v_{01}=v_{02}=q=0.
\eq
Same as case $1)$ except that ${\cal N}$ is given by eq. (7.14) with (7.17). 

\item 

\eql(7.39)
v_{01}=v_{02}=v_{12}=v_{21}=0.
\eq
\eql(7.40)
H_0=-v{\cal N},\;\; H_1=\sqrt{{u_0\over q_0}}{\cal A},\;\;
H_2=\sqrt{{u_0\over q_0}}{\cal B},\;\;\lambda={1\over q_0}.
\eq
${\cal A}$, ${\cal B}$ and ${\cal N}$ are given by eq. (7.24-5).

\item 
\eql(7.41)
v_{01}=v_{21}=0,\quad q=q_0={1\over\lambda}.
\eq
$$
H_0=-v{\cal N},\;\; H_1={u_0\over v_{12}}{\cal A},\;\;
H_2={v_{12}\over q_0}{\cal B},\;\;\lambda={1\over q_0}
$$
${\cal A}$, ${\cal B}$ and ${\cal N}$ are given by eq. (7.29) and ${\cal
N}$ by eq. (7.31) with (7.32).

\item
$$v_{02}=v_{01}=v_{21}=0,\quad q=0.$$
Everything as in case $4)$ except ${\cal N}$ is given by eq. (7.31) with
eq.(7.33).

\end{enumerate}

This closes the problem of Fock representations of the algebra (7.7).  We
now turn to the physical problem.  This implies using the definitions (5.19)
and (5.20) of the parameters $u_{\mu}$ and $v_{\mu\nu}$.  We first have to
give the boundary rates which give eq. (7.1).  Using (2.7) we have the
solution
\eql(7.42)
\begin{array}{rcl}
L_2^1&=&L_0^1=R_1^2=R_0^2=0\\
{\cal L}_1&=&L_1^0 L_0^2 + L_1^2 L_1^0 + L_1^2 L_2^0\\
{\cal R}_2&=&R^1_0 R^0_2 + R^1_2 R^0_1 + R^1_2 R^0_2
\end{array}
\eq

We now compute ${\cal B}_{\mu}$ and $x_{\mu}$ using eqs. 
(5.8) and (5.9):
\eql(7.43)
\begin{array}{rcl}
\delta_0&=&\xi(R_0^1 L_1^2 + L_0^2 R_0^1 + L_0^2 R_2^1)\\
\delta_1&=&\xi(R_1^0 L_0^2 + R_1^0 L_1^2 + R_2^0 L_1^2 + {\cal L}_1)\\
\delta_2&=&\xi(R_0^1 L_2^0 + R_2^1 L_2^0 + R_2^1 L_1^0 + {\cal R}_2)
\end{array}
\eq
and
\eql(7.44)
\begin{array}{rcl}
x_0&=&\xi(L_0^2{\cal R}_2 - R_0^1 {\cal L}_1)\\
x_1&=&\xi(L_1^2 {\cal R}_2 + (R_0^1 + R_2^1){\cal L}_1)\\
x_2&=&-\xi((L_0^2+L_1^2){\cal R}_2 + R_2^1 {\cal L}_1).
\end{array}
\eq
We now implement the conditions (7.35) to get
\eql(7.45)
y_0=0,\quad y_1=g_{10}(1-q),\quad y_2=-g_{02}(1-q).
\eq

We have to consider the $5$ cases separately.

\begin{enumerate}

\item From (7.37) and (5.20) we get  
\eql(7.46)
y_1=0=y_2.
\eq
Taking into account the positivity
of the rates,   the definition (5.17) and the relation (5.6) 
 this can be excluded.

\item From eq. (7.38) and (5.19) we have
\eql(7.47)
g_{20}=0=g_{01},\quad y_0=0,\;\; y_1=g_{10},\;\; y_2=-g_{02}.
\eq
Using the definitions (5.19) and (5.20) we get 
$$v_{10}=v_{20}=1$$
and
$$v_{12}=1-q_0+{g_{02}\over g_{21}},\;
v_{21}=1-q_0+{g_{10}\over g_{21}},\;
u_0=1-q_0+{g_{10}+g_{02}\over g_{21}},$$
and the following constraints on the rates coming from the definition (5.17)
of $y_{\mu}$ and from from eq. (7.47):
\eql(7.48)
\begin{array}{rcl}
L_0^2{\cal R}_2&=&R_0^1 {\cal L}_1\\
g_{10}&=&\dis {L_1^2 {\cal R}_2 + (R_0^1 + R_2^1){\cal L}_1\over
R_1^0 L_0^2 + R_1^0 L_1^2 + R_2^0 L_1^2 + {\cal L}_1}\\
g_{02}&=&\dis {(L_0^2+L_1^2){\cal R}_2 + R_2^1 {\cal L}_1
\over R_0^1 L_2^0 + R_2^1 L_2^0 + R_2^1 L_1^0 + {\cal R}_2}
\end{array}
\eq

\item As a consequence of eqs.(7.39) and (7.45) we find:
\eql(7.49)
g_{10}=g_{02},\;\; g_{01}=g_{20},\;\; g_{12}-g_{21}=g_{10}-g_{01},
\eq
\eql(7.50)
y_0=0,\quad y_1=-y_2=g_{10}-g_{01}
\eq
and
\eql(7.51)
v_{12}=0=v_{21},\quad u_0=q_0-1,\quad v_{10}=v_{20}=1-q.
\eq
From eq. (7.50) we find the following constraints on the rates:
\eql(7.52)
\begin{array}{rcl}
L_0^2{\cal R}_2&=&R_0^1 {\cal L}_1\\
\dis {L_1^2 {\cal R}_2 + (R_0^1 + R_2^1){\cal L}_1\over
R_1^0 L_0^2 + R_1^0 L_1^2 + R_2^0 L_1^2 + {\cal L}_1}&=&
\dis {(L_0^2+L_1^2){\cal R}_2 + R_2^1 {\cal L}_1
\over R_0^1 L_2^0 + R_2^1 L_2^0 + R_2^1 L_1^0 + {\cal R}_2}\\
&=&g_{10}-g_{01}.
\end{array}
\eq

\item No solutions if we want to maintain positivity of the rates.

\item We find
\eql(7.53)
g_{21}+g_{10}=g_{12}\quad g_{20}=0=g_{01},
\eq
\eql(7.54)
y_0=0,\quad y_1=g_{10},\quad y_2=-g_{02}
\eq
and
\eql(7.55)
v_{10}=v_{20}=1,\quad v_{12}={g_{02}-g_{10}\over g_{21}},\quad
u_0={g_{02}\over g_{21}}.
\eq
The conditions on the rates can be derived from eqs. (7.54) and (5.17).
Notice that this case is a specialization of case 3.  The representation
is different because $z_2$ defined in eq. (7.8) and used for case 3
diverges ($v_{21}=0$).

\end{enumerate}

In Appendix A we consider the case of a CP-invariant stochastic process
\cite{Evans-etal}.  In this problem, the index $"0"$ denotes vacancies,
$"1"$ is the index for particles and $"2"$ for anti-particles.  This
implies supplementary conditions on the rates.  If we are interested in the
case where the CP-symmetry is explicitly broken \cite{ArHei,AH} one can use
the solutions corresponding to the cases $2)$, $3)$ and $5)$ described
above.  In all three cases the $D_{\alpha}$'s have the form
\eql(7.56)
D_{\alpha}=x_{\alpha} E_{\alpha} \; (x_{\alpha}\neq 0), \;\; 
D_{\alpha}=E_{\alpha} \; (x_{\alpha}=0)
\eq
where $E_{\alpha}$, which can be extracted from the formulae given above 
and are given explicitly in ref. \cite{AHR}, depend only on the bulk
rates.  The boundary rates enter in the expressions of $x_{\alpha}$
and through the fact that they have to ``match'' the bulk rates (see
eqs. (7.48) and (7.52)).  This observation allows us to see the 
important parameters of the problem.

Before closing this section, let us note that the case where only the
cofactors ${\cal L}_0$ and ${\cal R}_0$ vanish can be treated in a similar
way.  In this case, the algebras satisfied by the ${\cal A}$ and 
${\cal B}$ matrices (see equations (7.9a) and (7.28)) contain 
${\cal A}^2$ and ${\cal B}^2$ terms as well.  Fock representations for this
case are also known \cite{EssRit}.

\setcounter{equation}{0}
\section{Lower rank boundary intensity matrices}

In sections 6 and 7 the Fock representations were defined  by four 
conditions (see eqs. (5.24), (6.4) or (7.2)).  The number of
conditions depends on the rank of the boundary intensity matrices.  
In principle, one should examine all the possibilities enumerated in 
eqs. (4.6$-$12) for both left and right intensity matrices.  This is
a long exercise.  The main point is that once either $L_{\mu}^{\nu}$
or $R_{\mu}^{\nu}$ or both have all the principal minors to be zero, the 
number of conditions is smaller and one has more freedom for the bulk rates.
We shall give only four examples.

\noindent (a).  ${\cal L}_1\neq 0$, all other cofactors zero.  

We take the following to be the boundary intensity matrices:
\eql(8.1)
L=\left(\begin{array}{ccc}
-L_1^0-L_2^0 & 0 & L_0^2 \\
L_1^0        & 0 & L_1^2 \\
L_2^0        & 0 & -L_1^2 - L_0^2
\end{array}\right)
\quad\mbox{and}\quad
R=\left(\begin{array}{ccc}
-R_1^0  &  R_0^1 & 0 \\
 R_1^0  & -R_0^1 & 0 \\
 0      &  0     & 0
\end{array}\right).
\eq
We set $\xi = 1$ in eqs. (5.8) and (5.9) to get 
\eql(8.2)
\begin{array}{rcl}
x_0&=&-R_0^1{\cal L}_1, \\
x_1&=&R_0^1{\cal L}_1, \\
x_2&=&0, \\
\delta_0&=&R_0^1(L_1^2+L_0^2),\\
\delta_1&=&R_1^0(L_1^2+L_0^2)+{\cal L}_1,\\
\delta_2&=&R_0^1 L_2^0.
\end{array}
\eq

We get several constraints on the various parameters $v_{ij}$ and 
$u_k$ that appear in the algebra (5.22) and are given in terms of the
bulk and boundary rates in (5.19-20).  From the condition $x_2=0$, 
and by sandwiching (5.22) between $\langle 0|$ and $|0\rangle$, we
obtain
$$v_{21}=u_0; \qquad v_{20}=u_1=0.$$
From $\langle 0| H_1 H_2 H_0 H_2|0\rangle$ we obtain
$$v_{10}=0$$
from which, on calculating $\langle 0| H_1 H_0 H_2|0\rangle$, we get
$q_1=q_0=q$.  The correlator $\langle 0| H_1 H_0 H_1 H_2|0\rangle$
gives 
$$q_2=1+{v_{01}\over u_0}(q-1),$$
so we end up with 
\eql(8.3) 
\begin{array}{ccccccc}
\dis {x_0\over \delta_0}&=&(g_{01}-g_{10})&=&(g_{02}-g_{20})
&=&(g_{21}-g_{12})\\
v_{12}&=&v_{02}&=&(1-q)&&
\end{array}
\eq

The algebra is of a
similar form as case c) of section 7 (eq. 7.26), apart from a 
relabelling of the indices and the presence of an additional constant 
term:
\eql(8.4)
\begin{array}{ccc}
q H_1 H_2 - H_2 H_1 &=& v_{21} H_2 + (1-q) H_1 + v_{21}\\
q H_2 H_0 - H_0 H_2 &=& (1-q) H_0\\
q_2 H_0 H_1 - H_1 H_0 &=& v_{01} H_0 + u_2
\end{array}
\eq
Introducing
\eql(8.5)
H_0={u_2\over v_{01}}{\cal A},\;\; H_1=v_{21} {\cal B}, \;\;
H_2=(q-1){\cal N}, \;\; \lambda=q_2^{-1}\;\;\hbox{and} \;\; 
\gamma = -{g_{02}\over g_{01}}
\eq
we get (compare with eq. (7.28)):
\eql(8.6)
\begin{array}{ccc}
{\cal A}{\cal B} - \lambda{\cal B}{\cal A} &=& \dis\gamma(1+{\cal A})\\
{\cal A}{\cal N} - q {\cal N}{\cal A} &=& {\cal A}\\
{\cal N}{\cal B} - q {\cal B}{\cal N} &=&\dis{\cal B}-{\cal N}+{1\over 1-q}.
\end{array}
\eq
Also, instead of four Fock conditions as in section 7, we have only three:
\eql(8.7)
\langle 0|{\cal A}=\langle 0|{\cal N}=0,\quad {\cal B}|0\rangle=
\alpha{\cal A}|0\rangle,
\eq
with
\eql(8.8)
\alpha=({g_{02}\delta_0\over x_0(\delta_1-\delta_0)})({\delta_0\over \delta_1}
{R_1^0\over R_0^1}).
\eq
We have not looked for explicit represenations of this algebra.

\noindent (b) All the cofactors are zero.

We consider the case when only the following four boundary rates 
do not vanish:
\eql(8.9)
L_2^0, L_0^2, R_0^1 \;\; \mbox{and} \;\; R_1^0.
\eq
We find from (5.9)
\eql(8.10)
x_{\mu}=0 \quad (\mu=0,1,2)
\eq
and the algebra is a known quadratic algebra (see section 3)
\eql(8.11)
q_0 D_1 D_2 = D_2 D_1, \;\; q_1 D_2 D_0 = D_0 D_2, \;\;
q_2 D_0 D_1 = D_1 D_0,
\eq
with only  two conditions:
\eql(8.12)
D_1|0\rangle={R_1^0\over R_0^1}D_0|0\rangle \;\;\mbox{and}\;\;
\langle 0|D_2 = {L_2^0\over L_0^2}\langle 0| D_0.
\eq
All vacuum expectation values of degree $L$ can be expressed in terms
of $\langle 0 | D_0^L |0\rangle$. The appearance of the polynomial algebras
in this new context is very interesting since, as shown in section 3, we
know plenty of them, and not just for the simple exclusion processes.  This
opens up the possibility of a new class of solutions for more general
processes with open ends.

\noindent (c) All the cofactors are zero, but the non-zero boundary rates are:

\eql(8.13)
L_1^0, L_2^0, R_0^1 \;\; \mbox{and} \;\; R_1^0.
\eq
In this case the algebra is also given by (8.11) and $x_{\mu} =0$, 
($\mu =0,1,2$), but the generators satisfy the conditions

\eql(8.14)
\langle 0|D_0 = 0 \;\; \mbox{and} \;\; 
D_1|0\rangle  = {R_1^0 \over R_0^1}D_0
|0\rangle  .
\eq
It is interesting to observe that no boundary condition appears for the 
operator $D_2$. Using (8.11) and (8.14) it is simple to see that the 
ground state will have a the single word $D_2^L$, which corresponds to 
the lattice filled by particle ``2". This can be expected physically from 
(8.13) since the species ``2'' is never removed from the chain at either 
end, unlike the species ``0'' and ``1'', and is created at the left
end with a certain rate $L_2^0\neq 0$.  Therefore,   
when the time goes to infinity the system 
will end up with only particles of species ``2''.

\noindent (d) All cofactors are zero, but the non-zero boundary rates are:

\eql(8.15)
L_1^0, L_2^0, R_1^0 \;\; \mbox{and} \;\; R_2^0.
\eq
The algebra is again given by (8.11) but the conditions are 

\eql(8.16)
\langle 0|D_0 = D_0 |0\rangle = 0,
\eq
and no conditions for $D_1$ and $D_2$. In this case we have $L$ degenerate 
ground states expressed in terms of the words 
$D_1^n D_2^{L-n}$ ($n = 1,2,\ldots,L$). This is expected physically from 
(8.15) since in the large time limit, we should expect no 
$``0''$ particles and the 
particles $``1''$ and $``2''$ are conserved separately.

\setcounter{equation}{0}

\section{Conclusions}

This was a long journey. We will first list the mathematical results 
which are independent of the physical applications. In section 3 we have given
a list of polynomial algebras defined by a set of homogenous quadratic 
relations (see for example eqs.(3.25-27)).
In sections 6 and 7 we have given Fock 
representations for the algebra defined by eq.(5.22). Those are a special
class of the inhomogenous quadratic algebras defined by eq.(1.24). The 
representations are one dimensional (see eq.(5.16)), 
two-dimensional for the ``generic" case described in section 6,  
or infinite 
dimensional (see the five special cases of the algebra (7.36)).
These 
algebras can be useful for physical applications other than for the
case of open chains considered in this paper.
For example we used them 
ourselves while studying stochastic processes on a ring \cite{AHR}.
As for the physical applications relevant to this paper, 
we have looked for steady states of stochastic processes defined 
by bulk and boundary intensity matrices
(see eq. (1.1)). The probability distribution of the steady 
states is given by the expression (1.13) for chains with open ends. 

In our approach the 
same expression without $|0\rangle$  is used for a left 
open end and a closed right  
end. In the case of closed chains we have to drop the 
$\langle 0|$ and $|0\rangle$ symbols. Polynomial algebras are 
defined by taking the $X$'s equal to zero 
in eq.(1.11). They exist only for special choices of the bulk rates. 
A given polynomial algebra can be used in three places: for closed ends 
(see section 3), for the left end open and the right end closed when 
supplementary restrictions come from the first equation in (1.12) with 
the $X$'s are taken to zero (see section 4) and even for the case of a chain 
with open ends (see the example in section 8, eqs.(8.11-12)). As far as we 
know it is the first time that the problem with one or two closed ends
is solved using algebraic methods.

Most of this paper is dedicated to the problem of the chain with open ends
where we have taken c-numbers for the $X$'s in eqs.(1.11) and (1.12). We 
have confined ourselves to the simpler problem of exclusion processes. 
In solving this problem a crucial point is the rank of the boundary 
intensity matrices since they dictate the number of conditions which define
the Fock representations. In the 3 state problem, they can be four (see
sections 6 and 7) three or two (see section 8). In the case of four conditions 
an important role is played by the number of cofactors of the boundary 
intensity matrices which vanish.  All in all we have given many examples
which can be used for physical applications.  It is obvious how the
present method may be generalized for four or more states problems.  
The case of periodic boundary conditions is considered in \cite{AHR}.  

After writing this paper, one can ask where lies the mathematical beauty.  
Probably in the simpler cases where polynomial algebras can be used 
and in the algebras described in section 7.  The problem
of open boundaries with two states depends on $5$ parameters; the one with
three states on $17$ parameters and there is a price to pay in order 
to find solutions.  In this paper we have never properly exploited
the Krebs-Sandow theorem (eqs. (1.12-15)) since we have always chosen
representations of the $X_{\alpha}$'s to be scalars or zero.  The reason
is that we wanted to have quadratic algebras that are understood, like the
polynomial algebras or those of type (1.24), and we did not find new ones.

We would like to make one last comment which has
to do with the connection between finding the ground states by algebraic
methods and the integrability of the Hamiltonian.  As is shown in section 
$3$ several exactly integrable Hamiltonians, connected with stochastic 
dynamics, can have their ground state wave function written in terms 
of words defined by algebraic relations. But in general, the existence
of an algebraic form of the ground 
state wave function is not a sufficient condition in order to ensure 
exact integrability.

\vskip 0.8in

\noindent{\large \bf Acknowledgements}

\vskip 12pt

The work of F.C.A. was supported in part by 
FAPESP and CNPQ - Brazil.  The work of S.D. was supported in
part by the EPSRC grant GRJ25758, and he would like to thank
Universit\"at Bonn for their hospitality.
V.R. would like to thank the hospitality of 
the Weizmann Institute, City University, the Federal University of 
S\~ao Carlos, the Institute for Theoretical Physics at the UCSB,
and especially of SISSA where this work was done, and
also the DAAD, the FAPESP,  the National Science Foundation under the 
Grant No. PHY94-07194 and the EC TMR Programme 
for financial support.   We would like to thank 
P.F. Arndt, L. Dabrowski, B. Derrida, T. Heinzel, K. Krebs, 
D.Mukamel, V. Pasquier, M. Scheunert and especially 
P.P. Martin for discussions.

\appendix
\setcounter{equation}{0}
\section{  The CP-invariant steady state}

It was recently shown that in steady states CP-invariance can be 
spontaneously broken \cite{Evans-etal} and that the 
phase diagram cannot be obtained by mean field methods \cite{AHR}.
This is not the right place to describe the physics of the problem
and the interested reader should have a look at the references.
The model where this invariance is spontaneously broken  
is a special case of the 
problem studied in section 7, where the rates we choose  
are CP invariant.
Charge conjugation (C) is defined by the operation 
$1\rightarrow 2$, $2\rightarrow 1$ and $0\rightarrow 0$, while the 
parity (P) operator corresponds to interchanging left and right.

The non-vanishing boundary rates are 
$$
\begin{array}{cc}
L_1^2=R_2^1, \;\; L_1^0=R_2^0,& L_2^0=R_1^0,\;\; L_0^2=R_0^1 \\
{\cal L}_1={\cal R}_2, & {\cal L}_0={\cal L}_2={\cal R}_0={\cal R}_1=0.
\end{array}
\eqno(A.1)
$$
The bulk rates are 
$$
g_{10}=g_{02},\quad g_{01}=g_{20},\quad q_1=q_2=q,
\eqno(A.2)
$$
and $g_{12}$ and $g_{21}$ are arbitrary.

In ref. \cite{Evans-etal} a special choice was taken:
$$
L_1^0=R_2^0=\alpha \quad \mbox{and} \quad L_0^2=R_0^1=\beta,
\eqno(A.3)
$$
with all other boundary rates set to zero.  The non-vanishing bulk
rates were taken to be $g_{12}=g_{10}=g_{02}$.  Here we consider the 
general case.

From equations (7.43), (7.44) and (5.17) 
and choosing $\xi^{-1}=(L_0^2+2L_1^2)$ we obtain
$$
\begin{array}{cc}
\delta_0=L_0^2,&\delta_1=\delta_2=(L_1^0+L_2^0)\\
y_0=0,&y_1=-y_2=\dis{L_1^0 L_0^2\over L_1^0+L_2^0}+L_1^2.
\end{array}
\eqno(A.4)
$$
We now use the results of section 7 (equations (7.43)-(7.55)).  From
eq. (7.45) and (A.4) we have
$$
{L_1^0 L_0^2\over L_1^0+L_2^0}+L_1^2=g_{10}-g_{01}.
\eqno(A.5)
$$

We now consider the cases (2), (3) and (5) separately.

\noindent Case (2), $g_{20}=g_{01}=0.$

We use equations (7.47) and (7.48) to get
$$
{L_1^0 L_0^2\over L_1^0+L_2^0}+L_1^2=g_{10},
\eqno(A.6)
$$
and 
$$
\begin{array}{cc}
v_{10}=v_{20}=1,&v_{01}=v_{02}=0,\\
\dis v_{12}=v_{21}={g_{10}+g_{21}-g_{12}\over g_{21}},
&\dis u_0={2g_{10}+g_{21}-g_{12}\over g_{21}}.
\end{array}
\eqno(A.7)
$$
Using eq. (5.21) it is easy to show that
$$
D_1=\left({L_1^0+L_2^0\over L_0^2}\right)(1+H_1),\quad
D_2=\left({L_1^0+L_2^0\over L_0^2}\right)(1+H_2).
\eqno(A.8)
$$
Introducing
$$\lambda={1\over q_0}={g_{21}\over g_{12}}\quad \mbox{and}\quad
\omega={g_{10}\over g_{12}}
\eqno(A.9)$$
we have
$$
H_1=H_2^T=\sqrt{2\omega+\lambda-1}\left(\begin{array}{cccc}
u_1&v_1&0&\ldots\\
0&u_2&v_2&\ldots\\
0&0&u_3&\ldots\\
\vdots&\vdots&\vdots&\ddots
\end{array}\right),\quad 
D_0=\left(\begin{array}{cccc}
1&0&0&\ldots\\
0&0&0&\ldots\\
0&0&0&\ldots\\
\vdots&\vdots&\vdots&\ddots
\end{array}\right)
\eqno(A.10)
$$
where
$$
\begin{array}{rcl}
v_n^2&=&\dis \{n\}_{\lambda}\left(1+{ (\omega+\lambda-1)^2
\over (2\omega-1+\lambda)}\{n-1\}_{\lambda}\right)\\
u_n&=&\dis {(\omega+\lambda-1)
\over \sqrt{2\omega-1+\lambda}}\{n-1\}_{\lambda}.
\end{array}
\eqno(A.11)
$$

Notice that the boundary conditions appear only in (A.6) and in the 
normalization factors of $D_1$ and $D_2$.  They do not change the 
physics of the problem which is governed by $\omega$ and $\lambda$.
In other words, taking only the rates (A.3) or the general case (A.1) 
does not change the physical results. 

In ref. \cite{Evans-etal} only the case $\lambda=0$ was considered.

\noindent Case (3) $g_{10}=g_{02}$, $g_{01}=g_{20}$ and $g_{12}-g_{21}=
g_{10}-g_{01}$.  

We make use of eqs.(7.49$-$52).
The bulk rates and the boundary rates are related by the constraint:
$$
g_{10}-g_{01}={L_1^0 L_0^2\over L_1^0+L_2^0}+L_1^2.
\eqno(A.12)
$$
We then have 
$$
\begin{array}{rclrcl}
D_1&=&\dis {L_1^0 + L_2^0\over L_0^2}(1+R)&D_2&=&D_1^T\\
R&=&\left(\begin{array}{cccc}
0&r_1&0&\ldots\\
0&0&r_2&\ldots\\
0&0&0&\ldots\\
\vdots&\vdots&\vdots&\ddots
\end{array}\right),&
D_0&=&\left(\begin{array}{cccc}
s_1&0&0&\ldots\\
0&s_2&0&\ldots\\
0&0&s_3&\ldots\\
\vdots&\vdots&\vdots&\ddots
\end{array}\right),
\end{array}
\eqno(A.13)$$
where
$$
r_k^2=1-\lambda^k\quad\mbox{and}\quad s_k=1-\{k-1\}_q
\eqno(A.14)$$
with
$$
\lambda={g_{21}\over g_{12}}\quad\mbox{and}\quad q={g_{20}\over g_{02}}=
{g_{01}\over g_{10}}.
\eqno(A.15)$$
Notice once more that the boundary terms are not essential and in this 
case as well, the conditions (A.3) capture the whole physics.

\noindent Case (5) (see eqs. (7.53$-$55)). 

Since $g_{02}=g_{10}$, we get $v_{12}=0$ (see eq. (7.55)) which 
together with (7.41) brings us back to case (3).  

In conclusion, considering the most general boundary rates (four 
instead of two) does not change the domain of applications of the
algebraic approach.  What we have shown however, is that it can be
applied for a larger choice of bulk rates \cite{AH}.

\end{document}